\documentclass[a4paper]{article}
\usepackage{epsfig}
\usepackage{graphicx}  
\usepackage{color}     

\usepackage{color}
\usepackage{lscape}
\date{\today}
 \normalsize

\newcommand{\bmat}{\left(\begin{array}}
\newcommand{\emat}{\end{array}\right)}
\newcommand{\be}{\begin{equation}}
\newcommand{\ee}{\end{equation}}
\newcommand{\bea}{\begin{eqnarray}}
\newcommand{\eea}{\end{eqnarray}}

\def\gtwid{\mathrel{\raise.3ex\hbox{$>$\kern-.75em\lower1ex\hbox{$\sim$}}}}
\def\ltwid{\mathrel{\raise.3ex\hbox{$<$\kern-.75em\lower1ex\hbox{$\sim$}}}}
\def\gev{{\rm \, Ge\kern-0.125em V}}
\def\tev{{\rm \, Te\kern-0.125em V}}

\def    \be            {\begin{equation}}
\def    \ee            {\end{equation}}
\def    \bea           {\begin{eqnarray}}
\def    \eea           {\end{eqnarray}}

\def\d{\delta}
\def\n{\nu}

\def\om{\omega}

\def\sig{\sigma}

\def\th{\theta}

\def\nn{\nonumber}
\def\d{\delta}
\def\D{\Delta}
\def\s{\sigma}
\def\r{\rho}
\def\t{\theta}

\def\me{\langle m\rangle_e}
\def\mee{\langle m\rangle_{ee}}
\begin{document}
\renewcommand{\thefootnote}{\fnsymbol{footnote}}
\vspace{.3cm}

\title{\Large\bf One vanishing minor in the neutrino mass matrix  }

\author
{ \it \bf  E. I. Lashin$^{1,2,3}$\thanks{elashin@ictp.it} and N.
Chamoun$^{4,5}$\thanks{nchamoun@hiast.edu.sy} ,
\\ \small$^1$ The Abdus Salam ICTP, P.O. Box 586, 34100 Trieste, Italy. \\
\small$^2$ Ain Shams University, Faculty of Science, Cairo 11566,
Egypt.
\\ \small$^3$ Department of physics and Astronomy, College
of Science, King Saud University, Riyadh, Saudi Arabia,\\
\small$^4$  Physics Department, HIAST, P.O.Box 31983, Damascus,
Syria,\\
\small$^5$  Centro Brasileiro de Pesquisas Fisicas, Rua Dr Xavier Sigaud 150, Rio de Janeiro, Brazil.  }

\maketitle

\begin{center}
\small{\bf Abstract}\\[3mm]
\end{center}
We study a specific texture of the neutrino mass
matrix, namely the models with one $2\times 2$ subdeterminant equal to zero.  We carry out a complete
phenomenological analysis with all possible relevant correlations. Every pattern of the six possible
ones is found able to accommodate the experimental data, with three
cases allowing also for non-invertible mass matrices. We present symmetry realizations
for all the models.

\vspace{1.1cm}{\bf Keywords}: Neutrino masses,
\vspace{1.1cm}{\bf PACS numbers}: 14.60.Pq; 11.30.Hv; 14.60.St
\begin{minipage}[h]{14.0cm}
\end{minipage}
\vskip 0.3cm \hrule \vskip 0.5cm
\section{Introduction}
Massive neutrinos and flavor mixing are the common ingredients in the interpretation of
the Super-Kamiokande \cite{SK} experiment on the solar and atmospheric neutrinos. In
the flavor basis which identifies the flavor eigenstates of the charged leptons with
their mass eigenstates, the (effective) neutrino mass matrix $M_\nu$ has nine free
parameters: three masses ($m_1$, $m_2$ and $m_3$), three mixing angles($\theta_x$,
$\theta_y$ and $\theta_z$) and three phases (two Majorana-type $\rho$, $\sigma$ and one
Dirac-type $\delta$). The values of the masses and the mixing angles are, somehow,
constrained by data  \cite{SNO,KM,K2K,CHOOZ}; whereas the phases are completely
unrestricted by current data.

Many viable zero-textures were studied. No three independent zeros-texture can
accommodate the data, whereas nine patterns of two independent zeros-texture, out of
fifteen possible, can do this \cite{FGM,Xing}). A specific model realizing any of the
possible six patterns of one zero-texture is constructed in \cite{Xing03b},however, it
led always to non-invertible mass matrices, some of which are compatible with current
data.

The seesaw mechanism relates $M_\nu$ to the Dirac neutrino mass matrix $M_D$ and the
Majorana mass matrix of the right handed singlet neutrinos $M_R$ through:
\bea
M_\n &=& M_D M_R^{-1} M_D^T,
\label{see-saw}
\eea
A zero in the inverted mass matrix $M_\nu ^{-1}$, when it exists, is related to a zero
in $M_R$, when $M_D$ is diagonal. In this respect, symmetry realizations of zeros in
$M_R$ \cite{Grimus} leading to zero-textures in $M_\nu ^{-1}$ were studied
in \cite{Lavoura}, and seven patterns of two zeros-texture in $M_\nu ^{-1}$ were showed
to be viable. However, these textures do not apply in the case of non-invertible
$M_\nu$ where the zeros in $M_R$ reflect themselves only as zero minors of $M_\nu$.
For this, the class of two independent zero minors-textures in $M_\nu$, irrespective of
being invertible or not, were investigated in \cite{LashinChamoun}. Seven patterns
could accommodate the data, with some viable textures allowing for singular mass matrix
with $m_3$ and $\th_z $ equal to zero.

One can generalize the zero-textures in $M_\nu$ in other ways. For instance, textures
of vanishing two-subtraces were tackled in \cite{SayedTrace} and eight patterns were
acceptable phenomenologically. In this paper, we apply the phenomenological analysis
of \cite{LashinChamoun,SayedTrace,dev} to study the textures of one vanishing minor in
$M_\nu$, without presupposing the invertibility of $M_\nu$, nor assuming any specific
model albeit we showed possible theoretical realizations of the patterns.

With the vanishing constraint (two real conditions) and the input of $\Delta m^2_{\mbox{sol}}$
fixed to its experimental central value, one still needs to know six parameters in order to
determine the mass matrix. We take the mixing and phase angles to be these parameters,
  and so we vary the values of the mixing angles $\theta_x, \theta_y$
and $\theta_z$ over their allowed experimental ranges whereas the unconstrained phase angles $\delta,\rho$
and $\sigma$ span their whole ranges. In this manner, one can obtain in the parameter space of
 $\theta_x, \theta_y, \theta_z, \delta, \rho$ and $\sigma $ the regions that are consistent with all
 other experimental constraints. Moreover, one can study the correlations between any two physical neutrino
 parameters $x$ and $y$ by plotting all the points $(x,y)$ corresponding to acceptable points in the parameter
 space. We found that all the six patterns, with three among them allowing for zero $m_3$, could
accommodate the data. Furthermore, four patterns exhibit all sorts of neutrino mass hierarchies, whereas
one pattern allows solely for an inverted hierarchy in contrast to another pattern showing only normal and
 degenerate hierarchies.

The plan of the paper is as follows: in section $2$, we review
the standard notation for the neutrino mass matrix
and its relation to the experimental constraints. In section $3$,
we present the texture of one vanishing minor in $M_\nu$ and find the corresponding expressions of
the two neutrino mass ratios. In section $4$, we classify all the patterns and present
the results and the phenomenological analysis of each case. We present symmetry
realizations of all models in section $5$ and end up with conclusions in section $6$.

\section{Standard notation}

In the flavor basis, which diagonalizes the charged lepton mass matrix, the symmetric
neutrino mass matrix $M_\nu$ can be diagonalized by a unitary transformation,
\begin{equation}
V^{\dagger} M_\nu\; V^{*} \; = \; \left (\matrix{ m_1 & 0 & 0 \cr 0 & m_2 & 0
\cr 0 & 0 & m_3 \cr} \right ), \;
\end{equation}
with $m_i$ (for $i=1,2,3$) real and positive. We introduce the mixing
angles $(\theta_x, \theta_y, \theta_z)$ and the phases ($\delta,\rho,\sigma$)
such that \cite{Xing}:
\bea
V &=& UP \\ P &=& \mbox{diag}(e^{i\rho},e^{i\sigma},1) \\ U \; &=& \;
\left ( \matrix{ c_x c_z & s_x c_z & s_z \cr - c_x s_y
s_z - s_x c_y e^{-i\delta} & - s_x s_y s_z + c_x c_y e^{-i\delta}
& s_y c_z \cr - c_x c_y s_z + s_x s_y e^{-i\delta} & - s_x c_y s_z
- c_x s_y e^{-i\delta} & c_y c_z \cr } \right ) \; ,
\eea
(with $s_x \equiv \sin\theta_x \ldots$).
We then have
\begin{equation}
M_\nu \; =\; U \left ( \matrix{ \lambda_1 & 0 & 0 \cr 0 &
\lambda_2 & 0 \cr 0 & 0 & \lambda_3 \cr} \right ) U^T. \;
\label{massdef}
\end{equation}
with
\begin{equation}
\lambda_1 \; =\; m_1 e^{2i\rho} \; , ~~~ \lambda_2 \; =\; m_2
e^{2i\sigma} \; , ~~~ \lambda_3 = m_3. \;
\end{equation}


In this parametrization, the mass matrix elements
are given by:
\bea
M_{\n\,11}&=& m_1 c_x^2 c_z^2 e^{2\,i\,\r} + m_2 s_x^2 c_z^2 e^{2\,i\,\s}
+ m_3\,s_z^2,\nn\\
M_{\n\,12}&=& m_1\left( - c_z s_z c_x^2 s_y e^{2\,i\,\r}
- c_z c_x s_x c_y e^{i\,(2\,\r-\d)}\right)
+ m_2\left( - c_z s_z s_x^2 s_y e^{2\,i\,\s}
+ c_z c_x s_x c_y e^{i\,(2\,\s-\d)}\right) + m_3 c_z s_z s_y,\nn\\
M_{\n\,13}&=& m_1\left( - c_z s_z c_x^2 s_y e^{2\,i\,\r}
+ c_z c_x s_x s_y e^{i\,(2\,\r-\d)}\right)
+ m_2\left( - c_z s_z s_x^2 c_y e^{2\,i\,\s}
- c_z c_x s_x s_y e^{i\,(2\,\s-\d)}\right) + m_3 c_z s_z c_y,\nn\\
M_{\n\,22}&=& m_1 \left( c_x s_z s_y  e^{i\,\r}
+ c_y s_x e^{i\,(\r-\d)}\right)^2 + m_2 \left( s_x s_z s_y  e^{i\,\s}
- c_y c_x e^{i\,(\s-\d)}\right)^2 + m_3 c_z^2 s_y^2, \nn\\
M_{\n\,33}&=& m_1 \left( c_x s_z c_y  e^{i\,\r}
- s_y s_x e^{i\,(\r-\d)}\right)^2 + m_2 \left( s_x s_z c_y  e^{i\,\s}
+ s_y c_x e^{i\,(\s-\d)}\right)^2 + m_3 c_z^2 c_y^2, \nn\\
M_{\n\, 23} &=& m_1\left( c_x^2 c_y s_y s_z^2  e^{2\,i\,\r}
 + s_z c_x s_x (c_y^2-s_y^2) e^{i\,(2\,\r-\d)} - c_y s_y s_x^2 e^{2\,i\,(\r-\d)}\right)
\nn\\
&& +  m_2\left( s_x^2 c_y s_y s_z^2  e^{2\,i\,\s}
 + s_z c_x s_x (s_y^2-c_y^2) e^{i\,(2\,\s-\d)} - c_y s_y c_x^2 e^{2\,i\,(\s-\d)}\right)
+ m_3 s_y c_y c_z^2.
\label{melements}
\eea
We see here that under the
transformation given by
\bea
T_1: &&\t_y\rightarrow {\pi \over 2} - \t_y\;\; \mbox{and}\;\; \d\rightarrow \d \pm \pi,
\label{sy1}
\eea
the mass matrix elements are transformed among themselves such that the indices $2$ and $3$ are swapped under
$T_1$ whereas the index $1$ remains invariant:
\bea
M_{\n 11} \leftrightarrow M_{\n 11}, && M_{\n 12} \leftrightarrow M_{\n 13}\nn\\
M_{\n 22} \leftrightarrow M_{\n 33}, && M_{\n 23} \leftrightarrow M_{\n23}.
\label{tr1}
\eea
There is another symmetry given by:
\bea
T_2: \r \rightarrow \pi - \r, & \s \rightarrow \pi - \s, & \d \rightarrow 2\,\pi-\d,
\label{sy2}
\eea
which changes the mass matrix into its complex conjugate i.e
\bea
M{_\n}_{ij}\left(T_2(\d,\r,\s)\right) = M^*_{\n ij} \left((\d,\r,\s)\right)
\eea

The above two symmetries $T_{1,2}$ are
very useful in classifying the models and in connecting the phenomenological analysis of patterns related
by these symmetries.

A remarkable merit of this parametrization is that its mixing
angles $(\theta_x, \theta_y, \theta_z)$ are directly related to
the mixing angles of solar, atmospheric and CHOOZ reactor neutrino
oscillations:
\begin{equation}
\theta_x \; \approx \; \theta_{\mbox{sol}} \; , ~~~~~ \theta_y \;
\approx \; \theta_{\mbox{atm}} \; , ~~~~~ \theta_z \; \approx \;
\theta_{\mbox{chz}}. \;
\end{equation}
Also we have,
\begin{equation}
\Delta m^2_{\mbox{sol}} \; = \;
\Delta m^2_{12} =m_2^2-m_1^2 \; , \;\Delta m^2_{\mbox{atm}}  \;
= \; |\Delta m^2_{23}|= \left|m_3^2-m_2^2\right|\;\; ,
\end{equation}
and the hierarchy of solar and
atmospheric neutrino mass-squared differences is characterized by
the parameter:
\begin{equation}
R_\nu \; \equiv \; \left | \frac{m^2_2 - m^2_1} {m^2_3 - m^2_2}
\right | \; \approx \; \frac{\Delta m^2_{\mbox{sol}}} {\Delta
m^2_{\mbox{atm}}} \; \ll \; 1 \; .
\end{equation}
Reactor nuclear experiments on beta-decay kinematics and
neutrinoless double-beta decay put constraints on the neutrino
mass scales characterized by the
effective electron-neutrino mass:
\begin{equation}
\langle
m\rangle_e \; = \; \sqrt{\sum_{i=1}^{3} \displaystyle \left (
|V_{e i}|^2 m^2_i \right )} \;\; ,
\end{equation}
and the effective Majorana mass term
$\langle m \rangle_{ee} $:
\begin{equation} \label{mee}
\langle m \rangle_{ee} \; = \; \left | m_1
V^2_{e1} + m_2 V^2_{e2} + m_3 V^2_{e3} \right | \; = \; \left | M_{\n 11} \right |^2.
\end{equation}
The Jarlskog rephasing invariant quantity \cite{jarlskog} that measure $CP$ violation in neutrino oscillation
is given by:
\begin{equation}\label{jg}
J = s_x\,c_x\,s_y\, c_y\, s_z\,c_z^2 \sin{\delta}
\end{equation}

Also, cosmological observations put an upper bound on the `sum'
parameter $\Sigma$:
\be
\Sigma = \sum_{i=1}^{3} m_i.
\ee

There are no experimental bounds on the phase angles, and we take the principal
value range for $\d, 2\r$ and $2 \s$ to be $[0,2\pi]$.
As to the other oscillation parameters, the experimental constraints give the following
 values with 1, 2, and 3-$\sigma$ errors \cite{fog,dev}:
\bea \label{Deltam}
\Delta m^2_{\mbox{atm}} &=& 2.6^{+0.2,\, 0.4,\,0.6}_{-0.2,\, 0.4,\,0.6} \times
10^{-3}\;\mbox{eV}^2,
\nonumber \\
\Delta m^2_{\mbox{sol}} &=& 7.9^{+0.3,\, 0.6,\, 1.0}_{-0.3,\, 0.6,\, 0.8}
\times 10^{-5}\;\mbox{eV}^2,
\eea
\bea
\sin^2\theta_{\mbox{atm}} = \left(
0.5^{+0.05,\,0.13,\, 0.18}_{-0.05,\, 0.12,\, 0.16}\right) &\longleftrightarrow & \th_y = \left(
45^{+3.44,\, 7.54,\, 10.55}_{-2.87,\,6.95,\,9.34}\right)\mbox{degree}, \nonumber \\
\sin^2\theta_{\mbox{sol}} = \left( 0.3^{+0.02,\, 0.06,\,0.10}_{-0.02,\,0.04,\,0.06}\right)
&\longleftrightarrow & \th_x = \left(
33.21^{+1.24,\,3.66,\,6.02}_{-1.27,\,2.56,\,3.88}\right)\mbox{degree}, \nonumber\\
\sin^2\theta_{\mbox{chz}} < 0.012,\, 0.025,\,0.040
&\longleftrightarrow & \th_z < \left(6.29,\, 9.10,\, 11.54\right)\mbox{degree}. \label{osdata}
\eea
The most stringent condition on any model required to fit
the data is the bound on the
$R_\n$ parameter:
\be
\label{Rconstraint}
R_\nu = \left(
0.0304^{+0.0038,\,0.0082,\,0.0141}_{-0.0033,\,0.0061,\,0.0082}\right).
\ee

Concerning the non oscillation parameters $\me,\, \Sigma$ and $\mee$,
we adopt the 2-$\sigma$ bounds for both $\me$ and $\Sigma$ as reported in \cite{fog}:
\bea
\langle m\rangle_e &<& 1.8\; \mbox{eV}, \nonumber \\
\Sigma &<& 1.4 \;\mbox{eV}.
\label{nosdata}
\eea
Due, in large, to the debate about the claimed observation of neutrinoless double beta decay, we left in our
phenomenological analysis
the effective Majorna mass term $\langle m\rangle_{ee}$ unconstrained.

\section{Neutrino mass matrices with one vanishing minor}

We denote by $C_{ij}$ the minor corresponding to the
$ij^{th}$ element (i.e. the determinant of the sub-matrix obtained
by deleting the $i^{th}$ row and the $j^{th}$ column of $M_\nu$). We have six
possibilities of having one minor vanishing. The vanishing minor condition is written
as:
\bea
\label{det1} M_{\nu\;ab}\;M_{\nu\;cd} - M_{\nu\;ij}\; M_{\nu\;mn} &=& 0,
\eea
then we have
\bea
\label{U's} \sum_{l,k=1}^{3}\left(
U_{al}U_{bl}U_{ck}U_{dk}-U_{il}U_{jl}U_{mk}U_{nk} \right)
\lambda_l \lambda_k &=& 0.
\eea
This leads to:
\bea
\frac{m_1}{m_3} &=&
\frac{Re(A_2 e^{-2i\sigma}) Im(A_1 e^{-2i\rho})-Re(A_1
e^{-2i\rho}) Im(A_2 e^{-2i\sigma}) }
{Re(A_3) Im(A_2 e^{-2i\sigma})-Re(A_2 e^{-2i\sigma}) Im(A_3)}  \\
\frac{m_2}{m_3} &=& \frac{Re(A_2 e^{-2i\sigma}) Im(A_1
e^{-2i\rho})-Re(A_1 e^{-2i\rho}) Im(A_2 e^{-2i\sigma})}{Re(A_1
e^{-2i\rho}) Im(A_3)-Re(A_3) Im(A_1 e^{-2i\rho})}
\eea
where
\bea
\label{Ah}
A_h&=& \left(
U_{al}U_{bl}U_{ck}U_{dk}-U_{il}U_{jl}U_{mk}U_{nk} \right) + \left(
l\leftrightarrow k \right),
\eea
with ($h,l,k$) are a
cyclic permutation of ($1,2,3$).

One can compute the analytical expressions, in terms of ($\theta_x, \theta_y, \theta_z,
\delta, \rho,$ and $\sigma$), of all the quantities measured experimentally. In order to explore the parameter
space of these six parameters, we have spanned the mixing angles, $\theta_x, \theta_y $ and $\theta_z$
over their experimentally allowed ranges
given in  eq.~\ref{osdata}, while the phases $\r, \s$ and $\delta$ were varied in their full ranges.
With $\Delta m_{\mbox{sol}}$ equal to its central value, we determined in the parameter space
 the acceptable regions compatible with the other experimental constraints given by
eqs.~\ref{Deltam}, \ref{Rconstraint} and \ref{nosdata}. One can then illustrate graphically all the possible correlations, in the three
 levels of $\s$-error, between any two physical neutrino parameters. We chose to plot for each pattern and for each
 type of hierarchy twenty six correlations at the 2$\s$ error level involving the parameters
 $(m_1,m_2, m_3,\t_x,\t_y,\t_z,\r,\s,\d,J,m_{ee})$ and the lowest neutrino mass (LNM). Moreover, for each
 parameter, one can determine the extremum values it can take according to the considered precision level, and we
 listed in tables these predictions for all the patterns and for the three $\s$-error levels.

We found that the resulting mass patterns could be classified into
three categories:
\begin{itemize}
\item Normal hierarchy: characterized by $m_1 < m_2 < m_3$ and
is denoted by ${\bf N}$. For this type of hierarchy, we imposed numerically the bound:
\be \frac{m_1}{m_3} < \frac{m_2}{m_3} < 0.7\ee
\item Inverted hierarchy: characterized
by $m_3 < m_1 < m_2$ and is denoted by ${\bf I}$. We imposed the corresponding bound:
\be \frac{m_2}{m_3} > \frac{m_1}{m_3} > 1.3\ee
\item Degenerate hierarchy: characterized
by $m_1\approx  m_2 \approx m_3$ and is denoted by ${\bf D}$. The corresponding numeric bound was
taken to be: \be 0.7 < \frac{m_1}{m_3} < \frac{m_2}{m_3} < 1.3\ee
\end{itemize}
Also, one should investigate the
possibility, for each pattern, to have singular (non-invertible) mass matrix. The
viable singular mass matrix is
characterized by one of the masses ($m_1,  m_2, \;
\mbox{and} \; m_3$) being equal to zero, as compatibility with the data prevents the
simultaneous vanishing of two masses and even vanishing of $m_2$ alone:
\begin{itemize}
\item The vanishing of $m_1$ implies that $A_1=0$ and the mass
spectrum of $m_2$ and $m_3$ takes the values $\sqrt{\Delta
m^2_{\mbox{sol}}}$ and $\sqrt{\Delta m^2_{\mbox{sol}}+\Delta m^2_{\mbox{atm}}}$
respectively.
\item The vanishing
of $m_3$ implies that $A_3=0$ and the mass spectrum of $m_2$
and $m_1$  takes the values $\sqrt{\Delta m^2_{\mbox{atm}}}$ and
$\sqrt{\Delta m^2_{\mbox{atm}}- \Delta m^2_{\mbox{sol}}}$
respectively.
\end{itemize}


The symmetry $T_1$ introduced in eqs.(\ref{sy1}-\ref{tr1}) induces
equivalence between different patterns of vanishing one minor as,
$C_{33} \leftrightarrow C_{22}$, and $ C_{31}\leftrightarrow C_{21}$. One should, however,
keep in mind that this equivalence for $\t_y$ is a reflection about the first bisectrix, i.e. it maps
the $\t_y$ from the first octant
to the second octant and vice versa. Similarly, the image points of the map differ in $\d$ from their
original points by a shift equal to $\pi$. This means that the accepted points for a pattern imply for the
equivalent pattern the same accepted points but after changing the $\t_y$ and $\d$ correspondingly.

Thus, it suffices now to present four possible cases, instead of six, corresponding
to one vanishing minor in $M_\nu$. Since the analytical expressions in terms of
($\theta_x ,\theta_y , \theta_z , \delta, \rho$ and $\sigma$) are quite complicated,
we state, simple writing permitting, only the leading terms of the expansions in powers
of $s_z$.

\section{Results of textures with one vanishing minor}
In this section, we shall present the results of our numerical analysis for the four possible independent
models based upon the approach described in the previous section. The coefficients $A's$ (eq.~\ref{Ah})
defining each model are presented. In order to get some interpretation of the numerical results, we present
also the analytical expressions of the mass ratios up to leading order
in $s_z$, except in the last pattern $C_{11}$ where we give the full, relatively simple, analytical
expressions of the mass ratios and other experimental parameters.

We organized the large number of correlation figures in plots, at the 2$\s$-error level, by dividing, where
applicable, each figure into left and right panels denoted accordingly by the letters L and R. Additional
labels (D,N and I) are attached to the plots to indicate the type of hierarchy (Degenerate, Normal and
Inverted, respectively). Any missing label D, N or I on the figures of
certain model would mean the absence of the corresponding hierarchy type in this model.

We listed in Tables (\ref{modelc33}) and (\ref{modelc32}), for the three types of hierarchy and the
three precision levels, the extremum values that the different parameters can take. The corresponding ranges should
get larger with higher-$\s$ precision levels. However, these bounds were evaluated by spanning the parameter space
with some given number (of order $(10^8-10^9)$) of points chosen randomly in the parameter
space. We found this way of random spanning more efficient than a regular meshing with nested loops. For a regular
meshing with a fixed step of `modest' order of 1 degree, we need around $10^{10}$ points to cover
the experimentally allowed space.
 However, in order to be efficient, the spanning needs a `dynamic' step for a finer meshing in the regions full of
  accepted points compensated by less spanning in the disallowed regions. With the random spanning we do not have this problem. Moreover, the randomness of our spanning allowed us to check the stability of
   our results for different randomly chosen points when we ran the programs several times. Thus, the values in the tables are meant to give only a strong qualitative indication. In particular, an
attainable zero value for $\th_z$ at one level implies this value is attainable for all higher $\s$ levels, even
though the corresponding values in the tables might be slightly larger than zero.

\subsection{ Pattern of vanishing minor $\mathbf {C_{33};\; M_{\nu 11}\,M_{\nu 22}- M_{\nu 12}\,M_{\nu 12}=0}$}
In this model, the relevant expressions for $A_1$, $A_2$ and $A_3$ are
\bea
 A_1 &=& (s_x s_y - s_z c_x c_y e^{-i\,\d})^2,\nn \\
 A_2 & = & (c_x s_y + s_z s_x c_y e^{-i\,\d})^2,\nn\\
 A_3 &=& c_z^2 c_y^2 e^{-2\,i\,\d},
 \label{minc33}
 \eea
leading to
\bea
 \frac{m_1}{m_3} &\approx& \frac{s_x^2  t_y^2s_{2\r - 2\s}}{s_{2\s - 2\d}} +
 O \left( s_z \right) \\
 \frac{m_2}{m_3} &\approx& \frac{c_x^2 t_y^2  s_{2\r - 2\s}}{s_{2\d - 2\r}}
  +O \left( s_z \right)
\eea

\begin{figure}[hbtp]
\centering
\begin{minipage}[l]{0.5\textwidth}
\epsfxsize=8.8cm
\centerline{\epsfbox{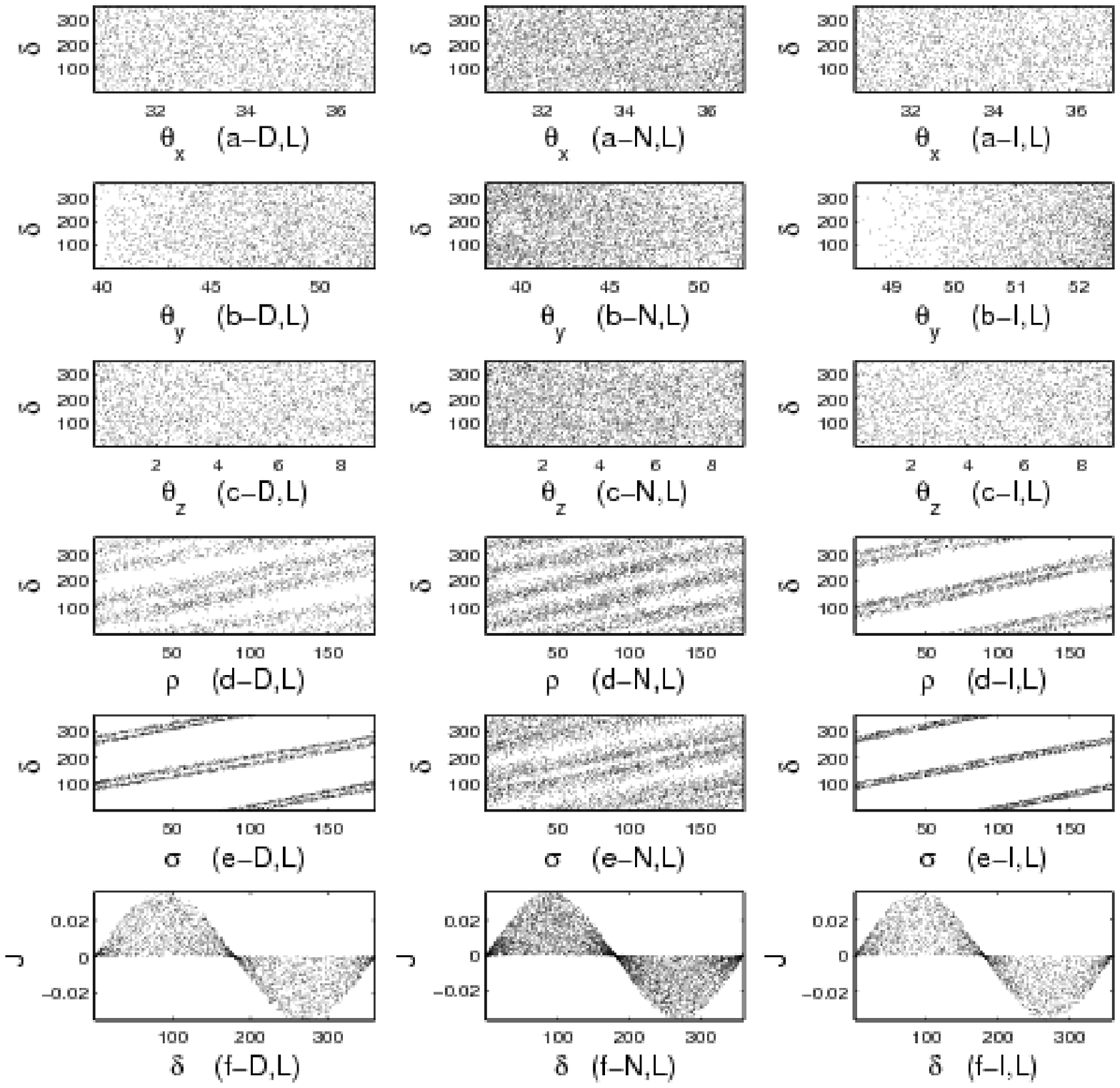}}
\end{minipage}%
\begin{minipage}[r]{0.5\textwidth}
\epsfxsize=8.8cm
\centerline{\epsfbox{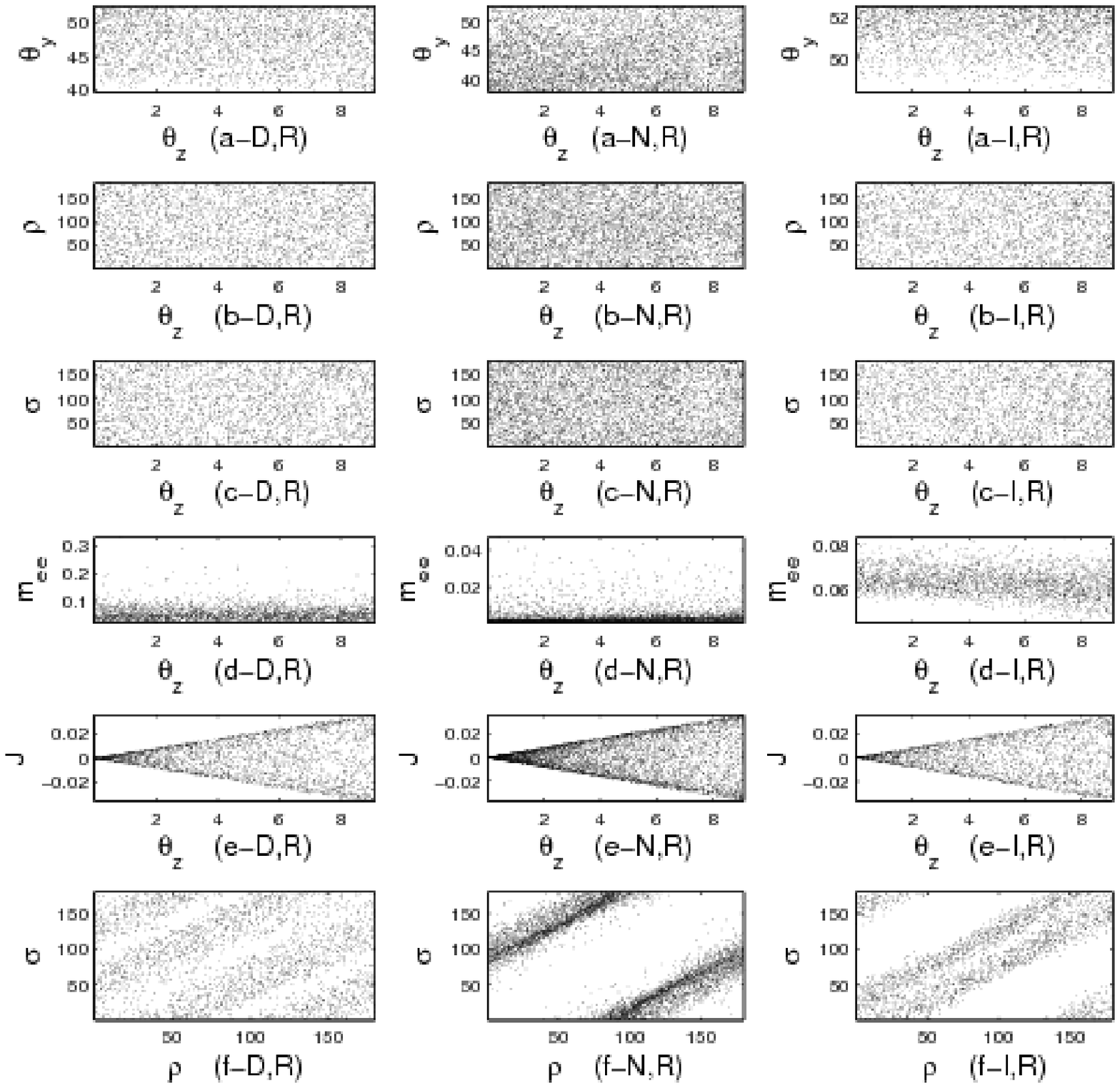}}
\end{minipage}
\vspace{0.5cm}
\caption{{\footnotesize Pattern $\mathbf C_{33}:$ Left panel presents correlations of $\delta$ against
mixing angles, CP-phases and $J$, while right panel shows the correlations of $\th_z$ against $\th_y$, $\r$ ,
$\s$, $m_{ee}$ and $J$, and also the correlation of $\r$ versus $\s$.}}
\label{c33fig1}
\end{figure}

In the left panel of Figure \ref{c33fig1}, we present the correlations of the angle $\d$ against the
mixing angles ($\t_x,\t_y,\t_z$), the CP phases ($\r,\s$) and the Jarlskog invariant quantity $J$, whereas
in the right panel we show the correlations of $\t_z$ against ($\t_y, \r, \s, J, \mee$), and the
correlation of $\r$ versus $\s$.

\begin{figure}[hbtp]
\centering
\begin{minipage}[l]{0.5\textwidth}
\epsfxsize=8.8cm
\centerline{\epsfbox{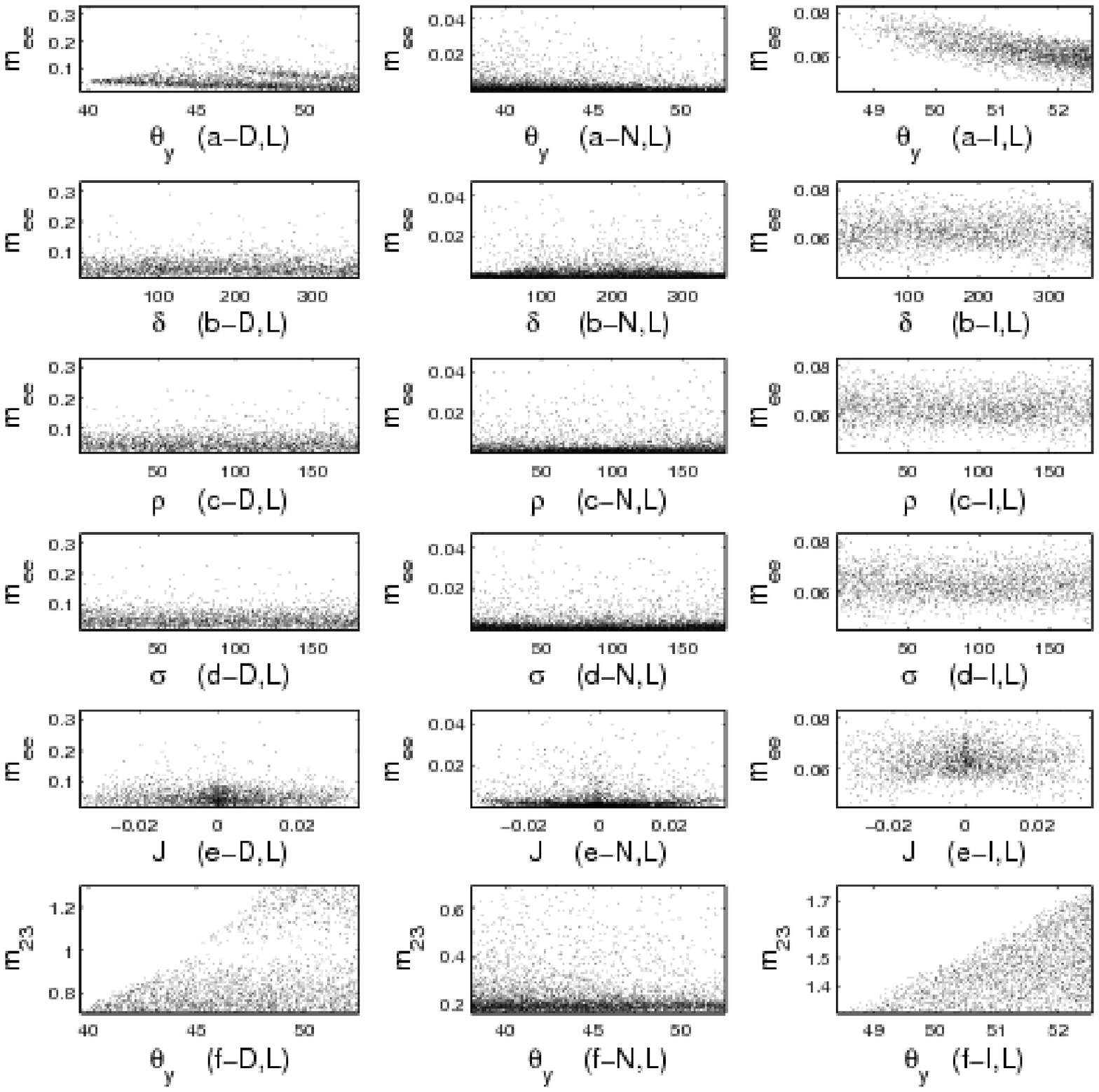}}
\end{minipage}%
\begin{minipage}[r]{0.5\textwidth}
\epsfxsize=8.8cm
\centerline{\epsfbox{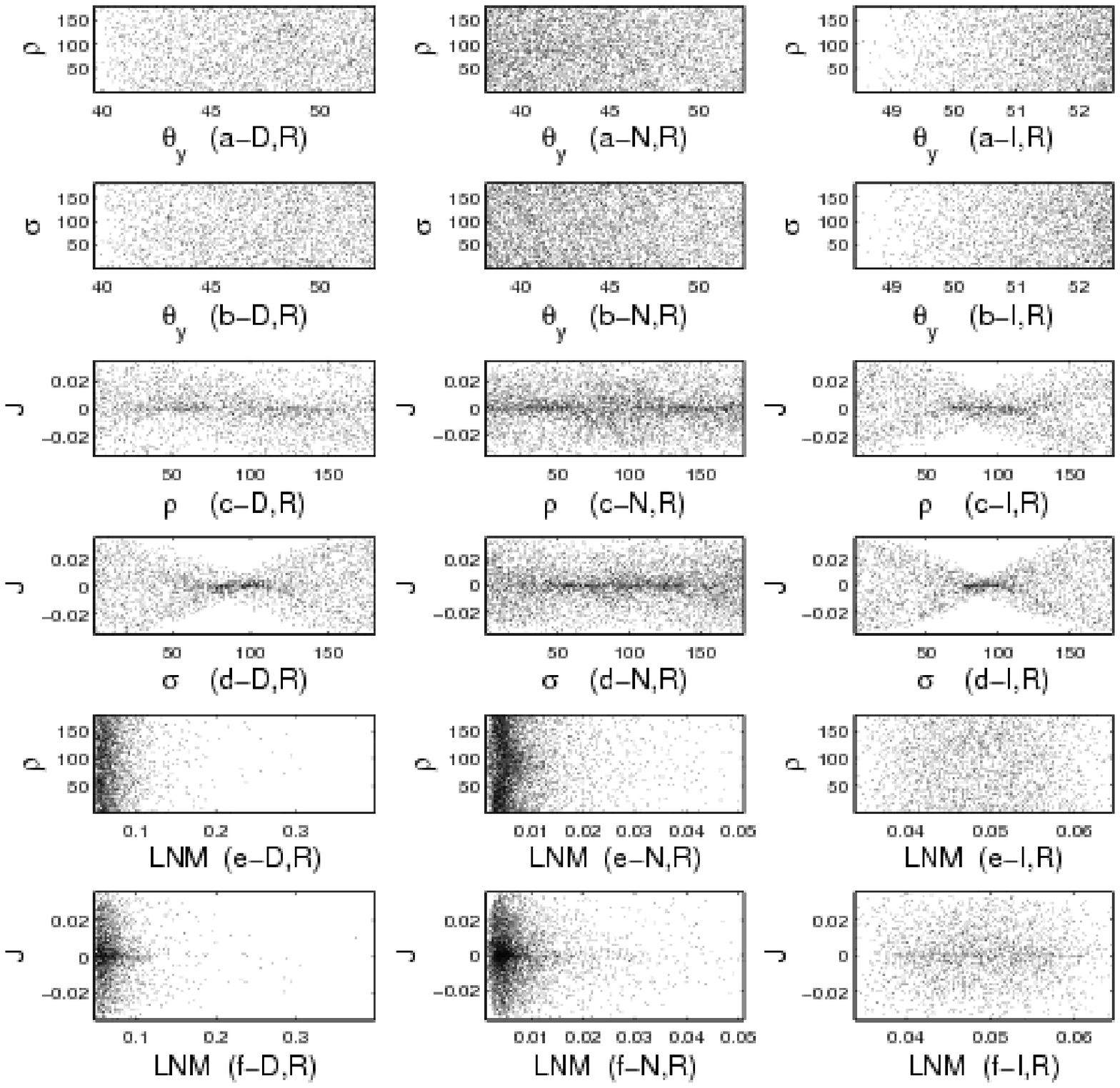}}
\end{minipage}
\vspace{0.5cm}
\caption{{\footnotesize Pattern $\mathbf C_{33}:$ Left panel presents correlations of $m_{ee}$ against
$\t_y$, $\d$,  $\r$ , $\s$, and $J$. It also shows the correlation between $m_2/m_3$ and $\t_y$.
 The right panel shows correlations of ($\r, \s$) against $\t_y$ and $J$ and those of the
  lowest neutrino mass (LNM) versus $\r$ and $J$.}}
\label{c33fig2}
\end{figure}

The left panel of Figure \ref{c33fig2} presents five correlations of $\mee$ against ($\t_y,\d, \r, \s$ and $J$)
and the correlation of $m_{23}={m_2 \over m_3}$ versus $\t_y$. As to the right panel
of this figure, it presents the  correlations of $(\r, \s)$ against $\t_y$ and $J$, and those
of $\mbox{LMN}$ versus $(\r, J)$.

\begin{figure}[hbtp]
\centering
\epsfxsize=10cm
\epsfbox{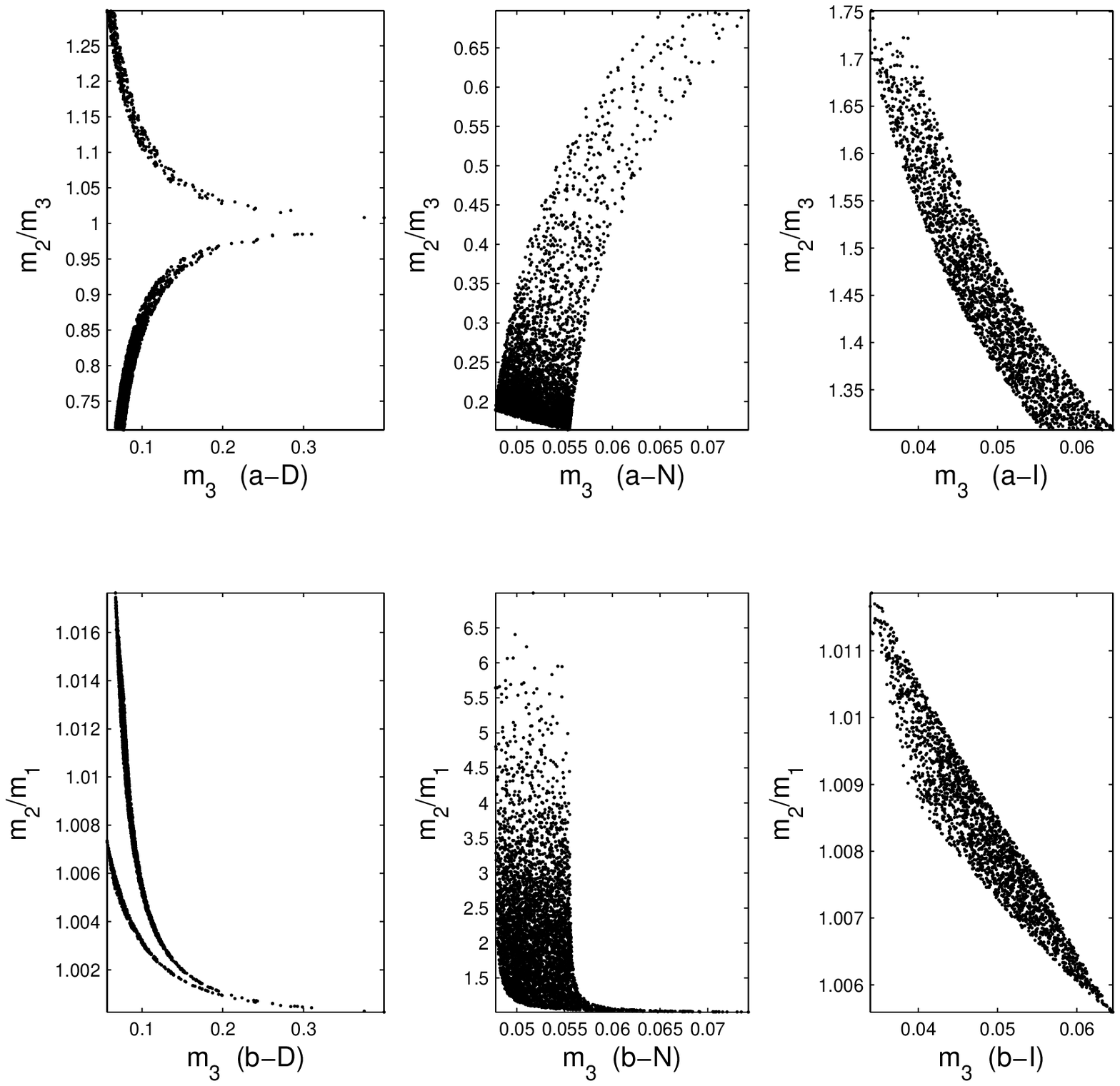}
\caption{{\footnotesize Pattern $\mathbf C_{33}:$ correlations of mass ratios ${m_2\over m_3}$ and
${m_2\over m_1}$ against $m_3$.}}
\label{c33fig3}
\end{figure}

As to Figure \ref{c33fig3}, and in a similar way, it presents two correlations of $m_3$ against $\frac{m_2}{m_3}$
and against $\frac{m_2}{m_1}$ for the three types of hierarchy. In all we have twenty six types of correlations for each hierarchy type.

We see in Figure~\ref{c33fig1}~(plots: a-L $\rightarrow$ c-L, a-R $\rightarrow$ c-R) that all the experimentally
allowed ranges of mixing angles, at $2 \s$
error levels, can be covered in this pattern except for inverted hierarchy type where $\t_y$ is restricted to be greater than $48^0$. This restriction on $\t_y$
distinguishes the inverted hierarchy type in this model. However, no obvious clear correlation can be revealed
in these plots. The plots (d-L, e-L) show that the phases are not constrained at all. However, in the case of
degenerate and inverted hierarchy, there is a strong linear correlation of $\d$ versus $\r$ and $\s$, whereas
 this correlation almost disappears in the normal
hierarchy case. There is also (plot f-R) a linear correlation between the Majorona phases which is
clearly apparent in the normal and inverted hierarchy types while a bit blurred in the degenerate case.

The correlations ($J,\d$) and ($J,\t_z$) have each a specific geometrical
shape which is hierarchy-type independent as it is clear from  Figure~\ref{c33fig1}~(plots: f-L ,e-R).
This behaviour can be understood from the formula of $J$ given in eq.~\ref{jg}. In fact, the correlation ($J,\d$)
can be seen as the superposition of many sinusoidal graphs in $\d$ whose `positive' amplitudes are determined by the
acceptable mixing angles, whereas the ($J,\t_z$) correlation is formed by the superposition of straight-lines
in $s_z \sim z$, for small $z$, whose slopes can be positive or negative depending on the sign of $s_\d$.

The correlations of $\mee$ against $(\t_x, \t_y, \d, \r, \s, J)$, as inferred from plot (d-R) of Figure~\ref{c33fig1}
 and from the left panel of
Figure~\ref{c33fig2} (plots: a-L $\rightarrow$ e-L), show that a lower bound for $\mee$ would generally constrain the allowed
parameter space. There is also a general tendency of decreasing $\mee$ with increasing $\t_y$ in the
case of inverted hierarchy (plot aI-L). Another important point concerning $\mee$ is that it can attain the zero-limit
in the normal hierarchy case, as is evident from the graphs or explicitly from the corresponding covered range in
Table~\ref{tab1}. This limit essentially corresponds to the case of vanishing $M_{\n 11}$ (equation \ref{mee})
which, when combined with vanishing
$C_{33}$ condition, implies vanishing $M_{\n 12}$. This means that in the limit of zero $m_{ee}$ we recover
 a corresponding two-zero
texture. It should be noted that this pattern of two-zero-entries texture is equivalent to the model
of vanishing two minors $C_{33}$ and $C_{32}$ \cite{LashinChamoun}. For the correlation of $m_2/m_3$ versus $\t_y$ (plot f-L)
we see that if the angle $\t_y$ is in the first octant then $m_2$ is less than $m_3$.

The right panel of Figure~\ref{c33fig2} does not show clear
correlation for $\t_y$ against $(\r, \s)$ (plots: a-R, b-R), whereas it indicates a correlation of  $J$ versus $(\r, \s)$
 (plots: c-R, d-R) which
is a direct consequence of the correlations of  $\d$ against $(\r, \s)$. The two correlations concerning the LNM
(plots: e-R, f-R) generally
reveal that as the LNM increases the parameter space becomes more restricted, and this seems to be a general
trend with increasing neutrino mass scale.

For the mass spectrum, as illustrated in Figure~\ref{c33fig3}, we see that the normal and inverted hierarchy are of
moderate type in that the ratios do not reach extremely high, nor low, values. The degenerate and inverted hierarchy
 types are characterized by nearly equal values of
$m_1$ and $m_2$. We also see that if $m_3$ is large enough then only the degenerate case with $m_1 \sim m_2$
can be compatible with data. We checked that there was no singular  texture which can accommodate the data, although
the limit $\t_z = 0$ can be reached. This can be seen from the coverable ranges of masses $m_1$ and $m_3$ in Table~\ref{tab1}.
This table also shows that no inverted hierarchy type of this pattern could be obtained at the 1$\s$ precision level.

\subsection{  Pattern of vanishing minor $\mathbf {C_{22};\; M_{\nu 11}\,M_{\nu 33}- M_{\nu 13}\,M_{\nu 13}=0}$}

In this model, the relevant expressions for $A_1$, $A_2$ and $A_3$
are
\bea
A_1 &=& (s_x  c_y + s_z c_x s_y e^{-i\,\d})^2,\nn \\
A_2 & = & (c_x c_y  - s_z s_x s_y e^{-i\,\d})^2,\nn\\
A_3 &=& c_z^2 s_y^2 e^{-2\,i\,\d},
\label{minc22}
\eea
We get
\bea
\frac{m_1}{m_3} &\approx& \frac{s_x^2 s_{2\r - 2\s}}{t_y^2 s_{2\s - 2\d}} +
O \left( s_z \right) \\
\frac{m_2}{m_3} &\approx& \frac{c_x^2  s_{2\r - 2\s}}{t_y^2 s_{2\d - 2\r}}
 +O \left( s_z \right)
 \eea

Again, there is no singular such texture which can accommodate the data. As for the plots, and since
this pattern is related by $T_2$-symmetry to the pattern $C_{33}$, they can be deduced from those of the
latter pattern but after changing $\t_y$ and $\d$ accordingly.

\subsection{  Pattern of vanishing minor $\mathbf {C_{31};\; M_{\nu 12}\,M_{\nu 23}- M_{\nu 13}\,M_{\nu 22}=0}$}
The relevant expressions for $A_1$, $A_2$ and $A_3$ for this model
are
\bea
A_1 &=& c_z c_x (s_y s_x - s_z c_x c_y\, e^{-i\,\d})\,e^{-i\,\d}\nn \\
A_2 & = & -c_z s_x (s_y c_x+s_z s_x c_y\, e^{-i\,\d})\,e^{-i\,\d},\nn\\
A_3 &=& s_z c_z c_y e^{-2\,i\,\d},
\label{minc13}
\eea
We obtain
\bea
\frac{m_1}{m_3} &\approx& \frac{t_y c_x s_x s_{2\r - 2\s}}{ s_{2\s - \d}\, s_z} +
O \left( s_z \right) \\
\frac{m_2}{m_3} &\approx& \frac{t_y c_x s_x s_{2\s - 2\r}}{s_{\d - 2\r}\,s_z}
 +O \left( s_z \right)
 \eea

\begin{figure}[hbtp]
\centering
\begin{minipage}[l]{0.5\textwidth}
\epsfxsize=8.8cm
\centerline{\epsfbox{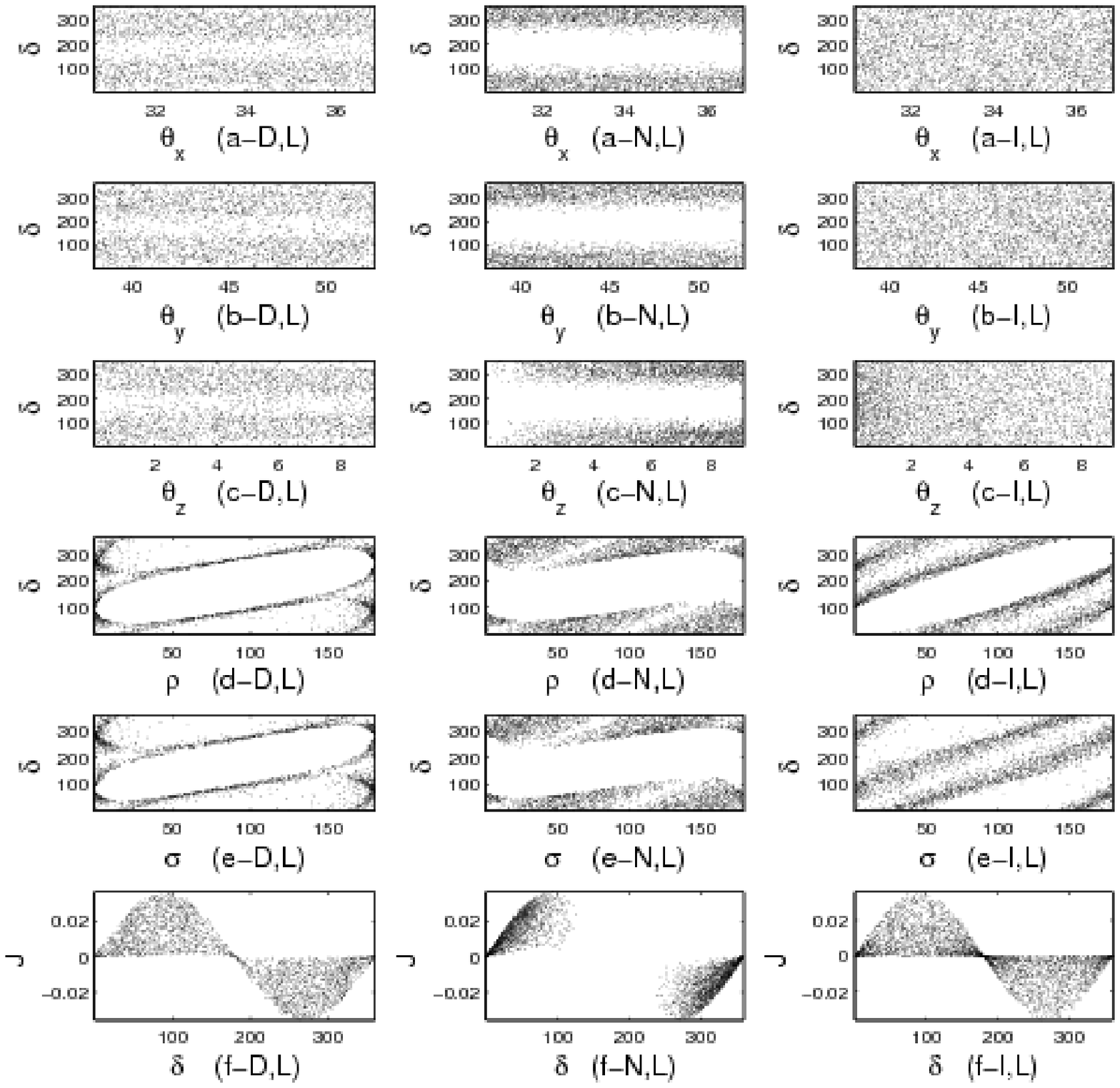}}
\end{minipage}%
\begin{minipage}[r]{0.5\textwidth}
\epsfxsize=8.8cm
\centerline{\epsfbox{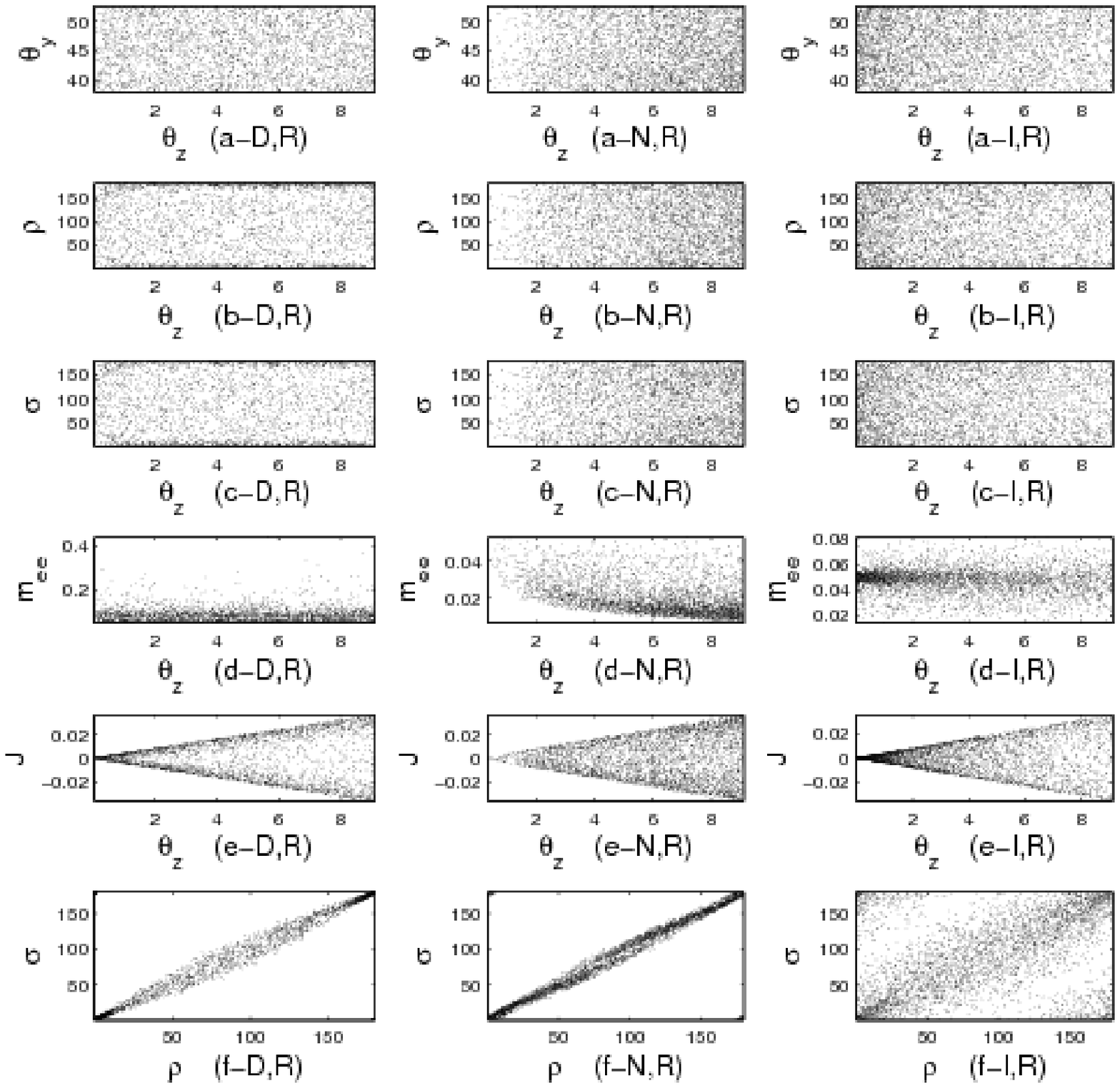}}
\end{minipage}
\vspace{0.5cm}
\caption{{\footnotesize Pattern $\mathbf C_{31}:$ Left panel presents correlations of $\delta$ against
mixing angles, CP-phases and $J$, while right panel shows the correlations of $\th_z$ against $\th_y$, $\r$ ,
$\s$, $m_{ee}$ and $J$, and also the correlation of $\r$ versus $\s$.}}
\label{c31fig1}
\end{figure}

\begin{figure}[hbtp]
\centering
\begin{minipage}[l]{0.5\textwidth}
\epsfxsize=8.8cm
\centerline{\epsfbox{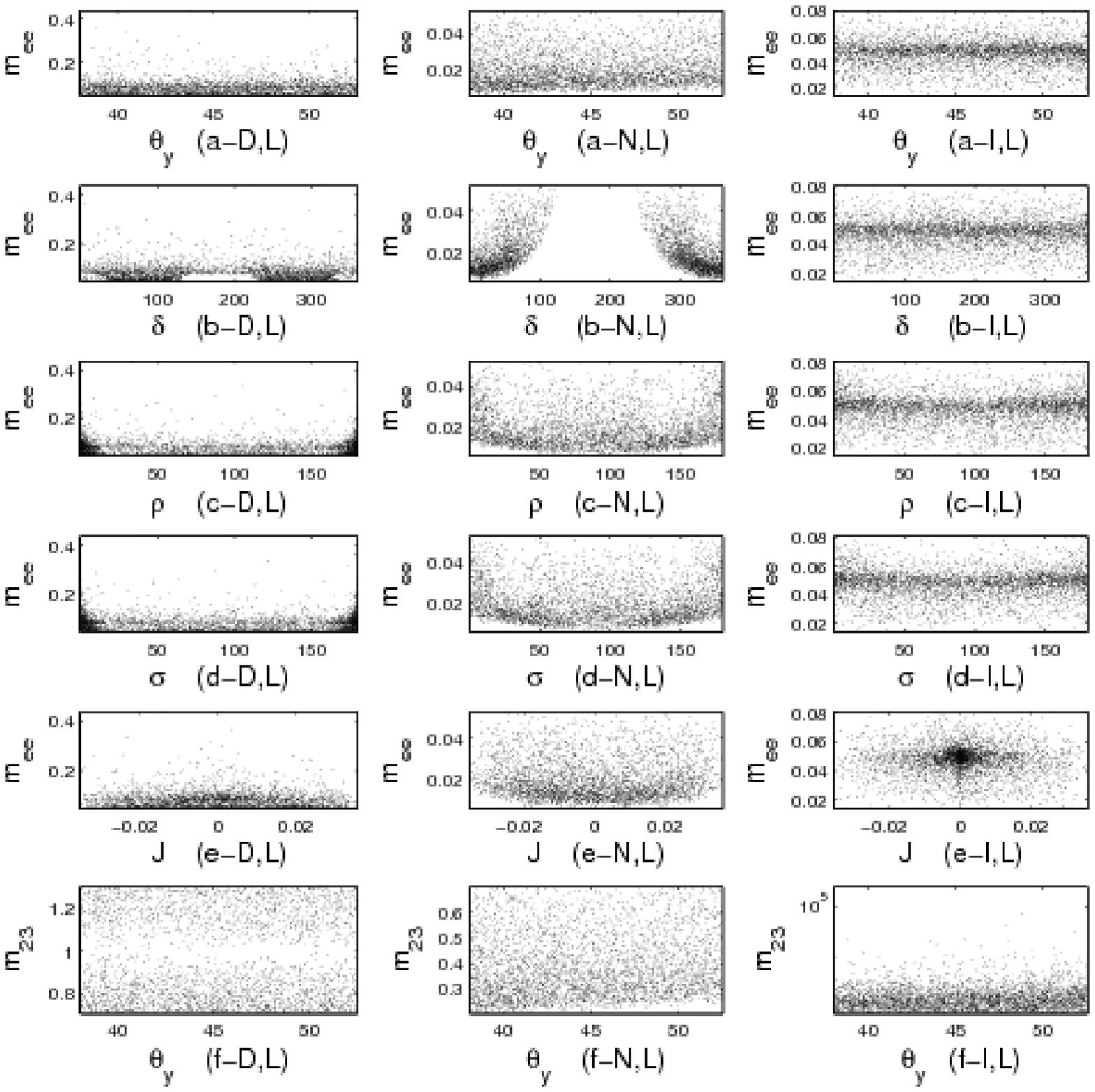}}
\end{minipage}%
\begin{minipage}[r]{0.5\textwidth}
\epsfxsize=8.8cm
\centerline{\epsfbox{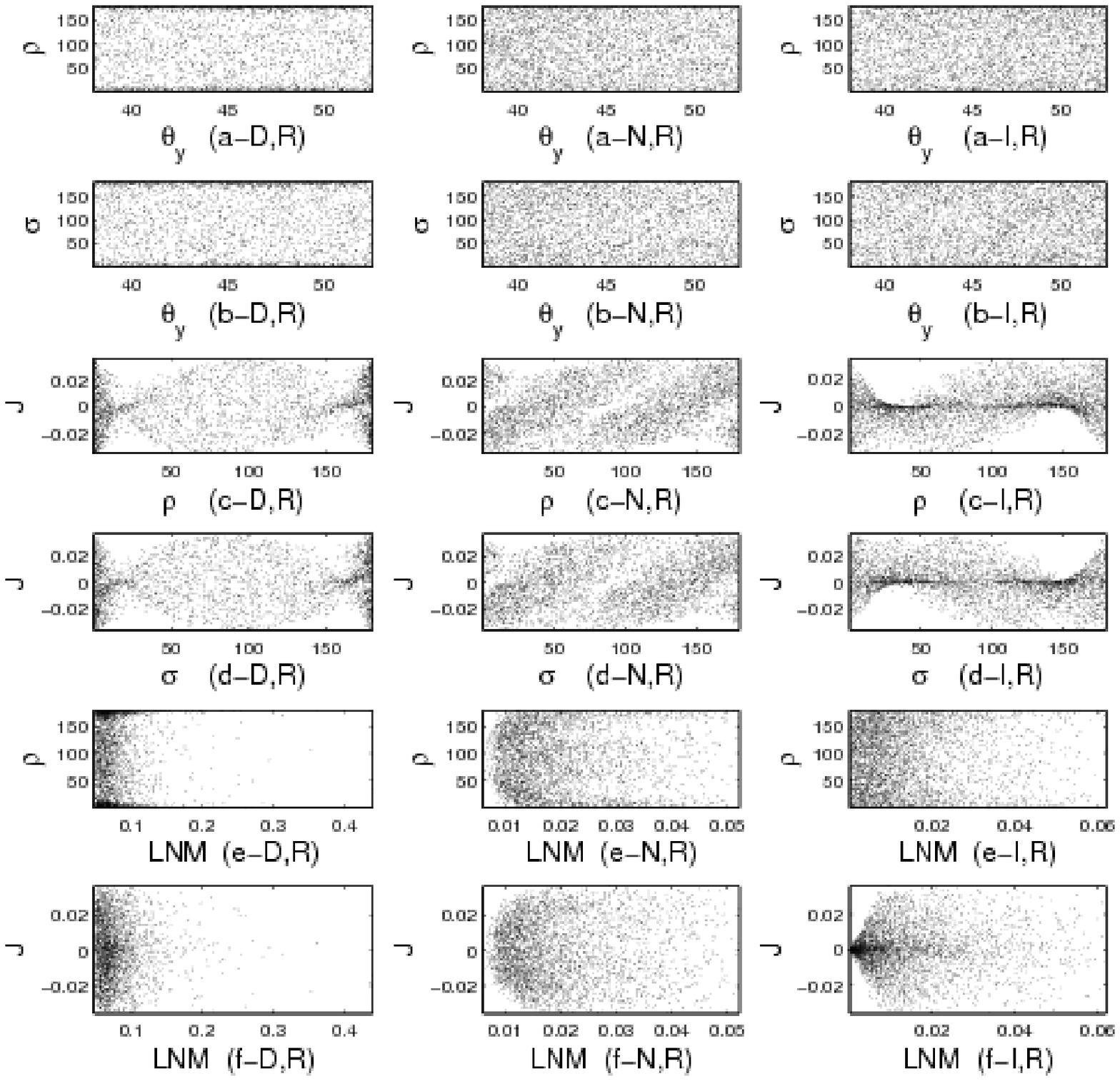}}
\end{minipage}
\vspace{0.5cm}
\caption{{\footnotesize Pattern $\mathbf C_{31}:$ Left panel presents correlations of $m_{ee}$ against
$\t_y$, $\d$,  $\r$ , $\s$, and $J$. It also shows the correlation between $m_2/m_3$ and $\t_y$.
 The right panel shows correlations of ($\r, \s$) against $\t_y$ and $J$ and those of the
  lowest neutrino mass (LNM) versus $\r$ and $J$.}}
\label{c31fig2}
\end{figure}

\begin{figure}[hbtp]
\centering
\epsfxsize=10cm
\epsfbox{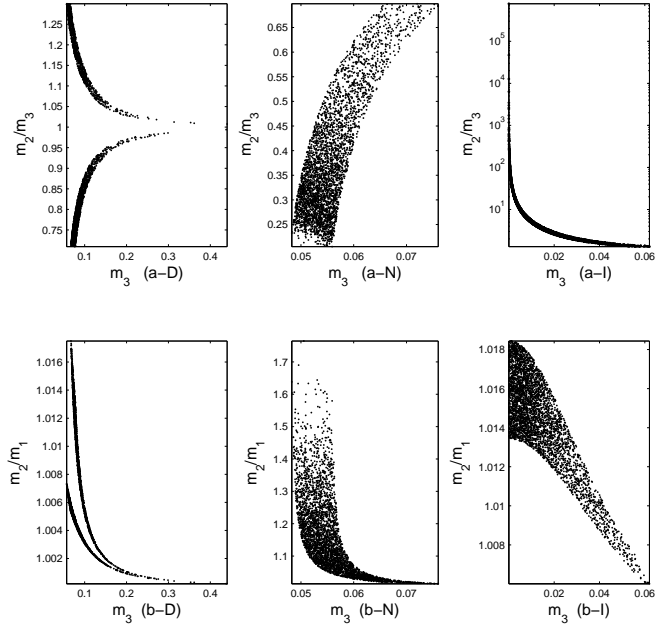}
\caption{{\footnotesize Pattern $\mathbf C_{31}:$ correlations of mass ratios ${m_2\over m_3}$ and
${m_2\over m_1}$ against $m_3$.}}
\label{c31fig3}
\end{figure}

We have also \bea R_\nu &=& \frac{c_{2\s} c_{2\s +2\d}-c_{2\r} c_{2\r -2\d}}{c_\d^2-c_{2\s}c_{2\s -2\d}}+O \left(
s_z
 \right)
\eea
We plot the correlations in Figures (\ref{c31fig1}, \ref{c31fig2}  and \ref{c31fig3}) with the same conventions as in
the case of $C_{33}$ pattern. Compared to the latter case, we see that the mixing angles ($\t_x,\t_y,\t_z$) can cover
all their allowable regions (Fig~\ref{c31fig1}, plots: a-L $\rightarrow$ c-L, a-R $\rightarrow$ c-R)
and in all hierarchy types. The linear
correlations of $\d$ versus $\r$ and $\s$ disappear in the inverted case, whereas they are replaced by Lissajous-like patterns
in the degenerate case
 (Fig~\ref{c31fig1}, plots: d-L, e-L). However,
there is an acute linear correlation between $\r$ and $\s$ (Fig~\ref{c31fig1}, plot: f-R) in the degenerate and normal cases.
The special `sinusoidal' and `isosceles' shapes of $J$ versus $\d$ and $\t_z$ remain
(Fig~\ref{c31fig1}, plots: f-L, e-R), but we note that in the normal case
the sinusoidal shape is concentrated for $\d$ in the first and fourth quarters, which would single out a disallowed
region for $\d$ ranging from $123^0$ to $242^0$ approximately. Again no clear correlation
involves $m_{ee}$ (Fig~\ref{c31fig1} plot d-R, Fig~\ref{c31fig2} plots: a-L $\rightarrow$ e-L). However, setting a lower bound on this parameter would constrain the parameter space only
in the degenerate case. Apart from the usual correlations of $J$ versus $\r$ and $\s$
(Fig~\ref{c31fig2} plots: c-R, d-R), originating from the correlation of $\d$ with $\r$ and $\s$, the other plots
doe not show clear correlations. We see from Table~\ref{tab1} that the limit $m_{ee} = 0$ is not attainable in this pattern.

For the mass spectrum, the plot b-I in Fig~\ref{c31fig3} tells us that the experimental data can be
accommodated in the inverted hierarchy type only when the two masses $m_1$ and $m_2$ are approximately equal. However, the
mass ratio-parameter $m_2/m_3$ (plot a-I) indicates a strong hierarchy. This is to be contrasted with the
normal type hierarchy case (plots a-N and b-N) where the hierarchy is mild and the mass ratios are of order $O(1)$.
We see also that in contrast to the pattern $C_{33}$, the limit $m_3 = 0$ can be reached.
In fact, there is a non-invertible such texture which can accommodate the current data, and this
happens only when $\t_z=0$ leading to $m_3=0$.

\subsection{  Pattern of vanishing minor $\mathbf {C_{32};\; M_{\nu 11}\,M_{\nu 23}- M_{\nu 21}\,M_{\nu 13}=0}$}

The relevant expressions for $A_1$, $A_2$ and $A_3$ for this model
are
\bea
A_1 &=& -\left(s_z c_x c_y e^{-i\,\d}-s_x s_y\right)
\left(s_x c_y  + s_z c_x s_y e^{-i\,\d}\right),\nn \\
A_2 & = & -\left(c_y s_z s_x e^{-i\,\d}+ s_y c_x \right)
\left(s_y s_z s_x e^{-i\,\d}-c_y c_x\right),\nn\\
A_3 &=& -c_z^2 s_y c_y e^{-2\,i\,\d}.
\label{minc32}
\eea
We get
\bea
\frac{m_1}{m_3} &\approx& \frac{s_x^2 s_{2\s - 2\r}}{ s_{2\s - 2\d}} +
O \left( s_z \right) \\
\frac{m_2}{m_3} &\approx& \frac{c_x^2  s_{2\r - 2\s}}{s_{2\r - 2\d}}
 +O \left( s_z \right)
 \eea

\begin{figure}[hbtp]
\centering
\begin{minipage}[l]{0.5\textwidth}
\epsfxsize=8.8cm
\centerline{\epsfbox{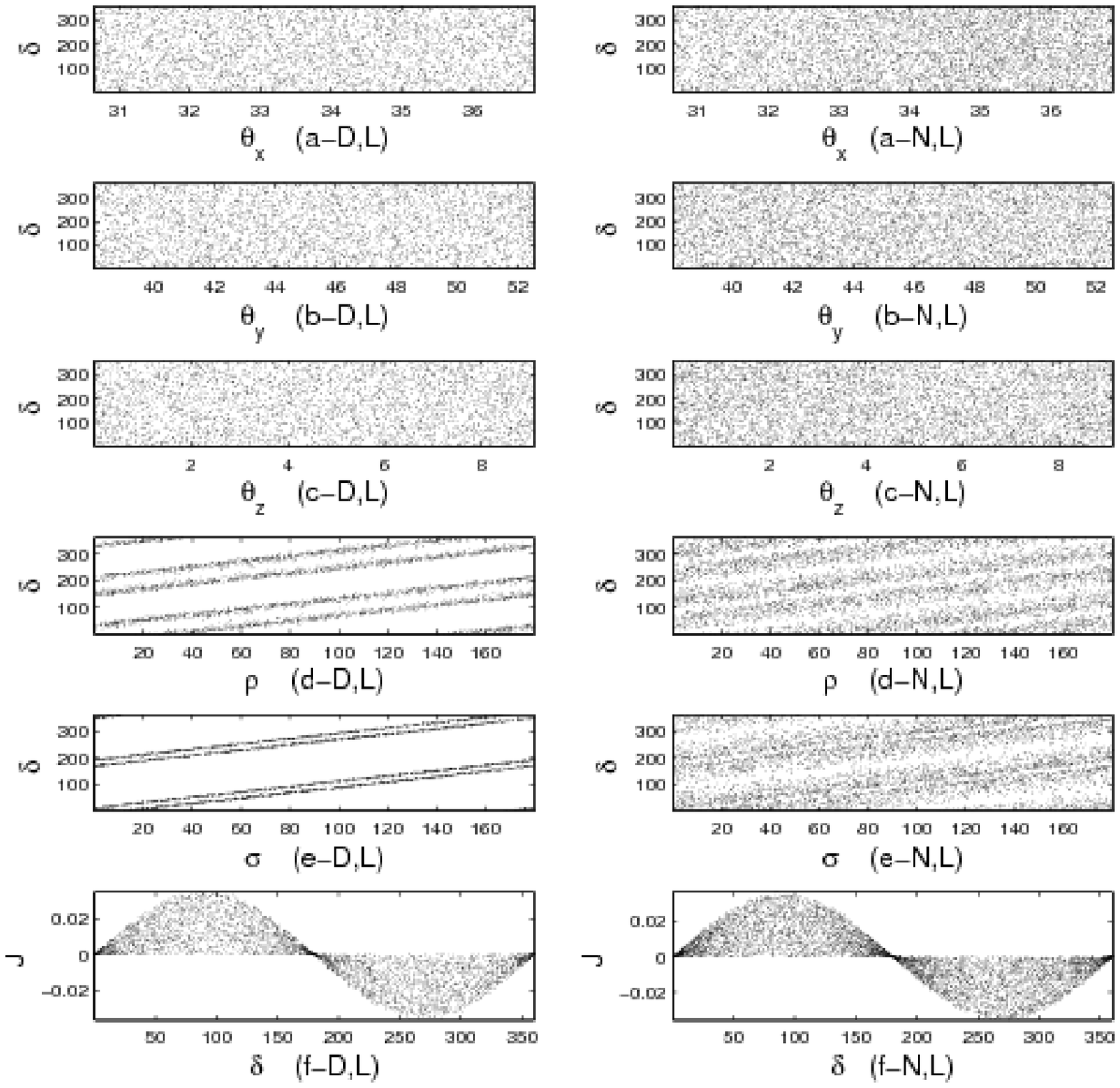}}
\end{minipage}%
\begin{minipage}[r]{0.5\textwidth}
\epsfxsize=8.8cm
\centerline{\epsfbox{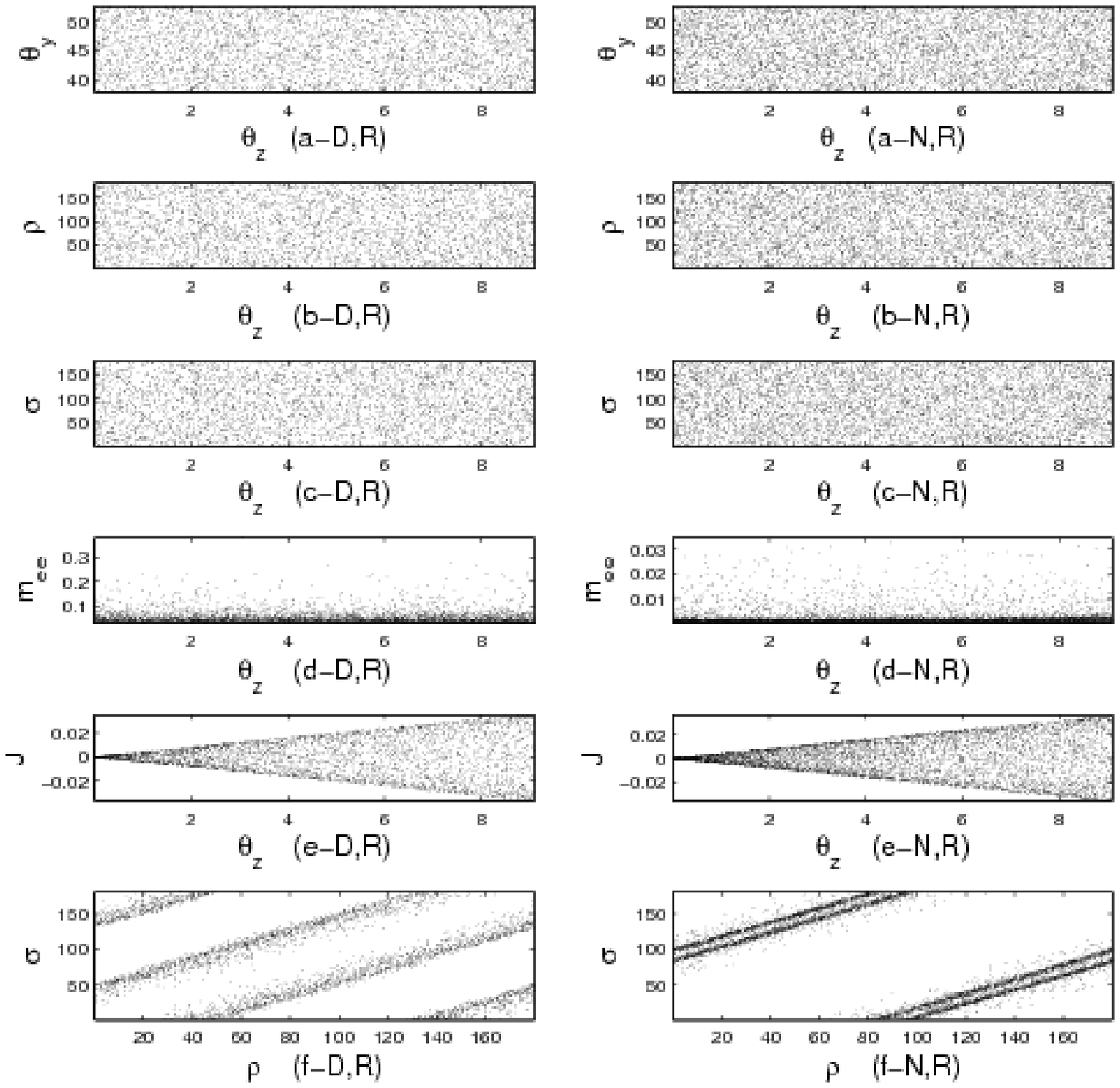}}
\end{minipage}
\vspace{0.5cm}
\caption{{\footnotesize Pattern $\mathbf C_{32}:$ Left panel presents correlations of $\delta$ against
mixing angles, CP-phases and $J$, while right panel shows the correlations of $\th_z$ against $\th_y$, $\r$ ,
$\s$, $m_{ee}$ and $J$, and also the correlation of $\r$ versus $\s$.}}
\label{c32fig1}
\end{figure}

\begin{figure}[hbtp]
\centering
\begin{minipage}[l]{0.5\textwidth}
\epsfxsize=8.8cm
\centerline{\epsfbox{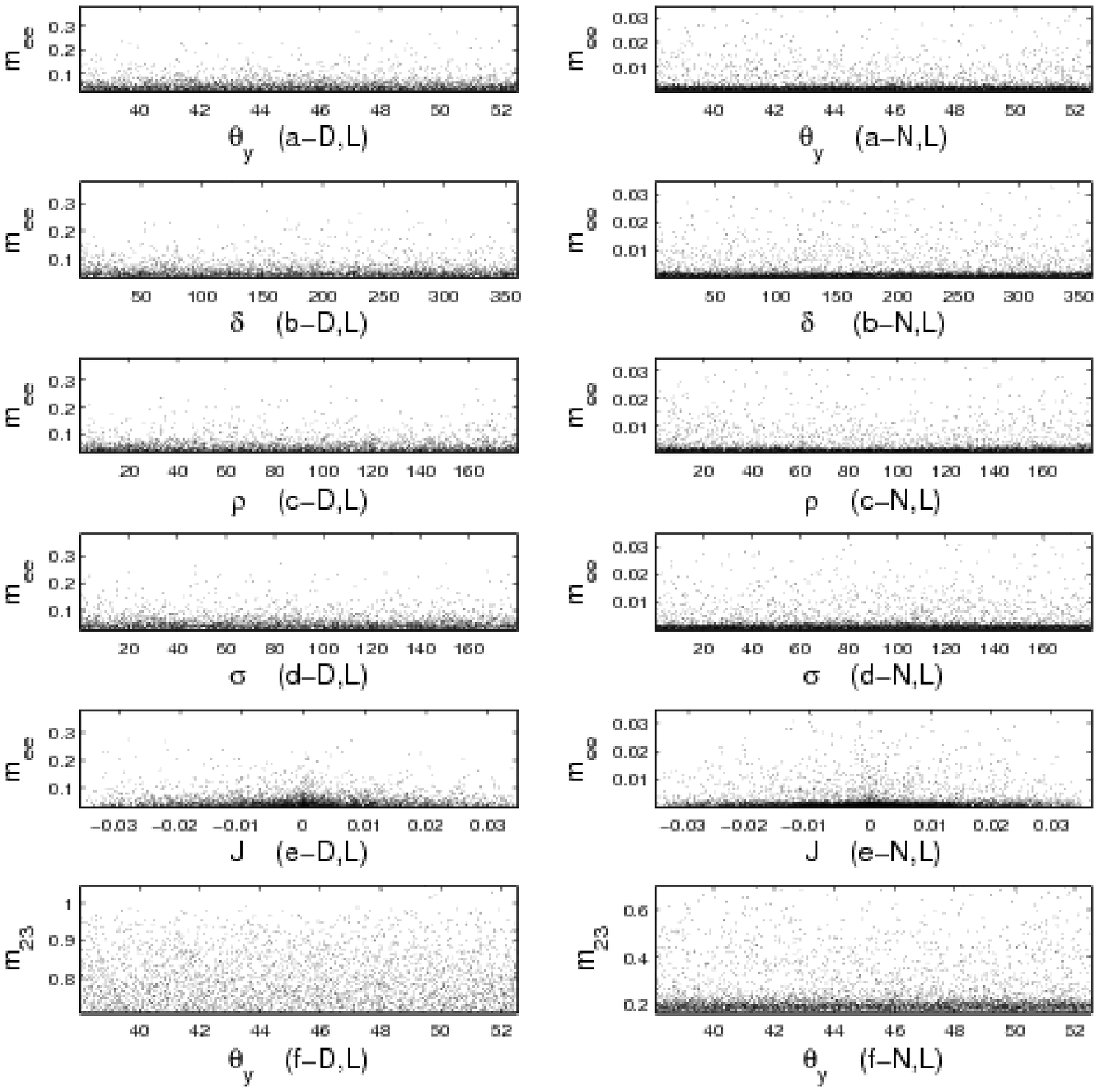}}
\end{minipage}%
\begin{minipage}[r]{0.5\textwidth}
\epsfxsize=8.8cm
\centerline{\epsfbox{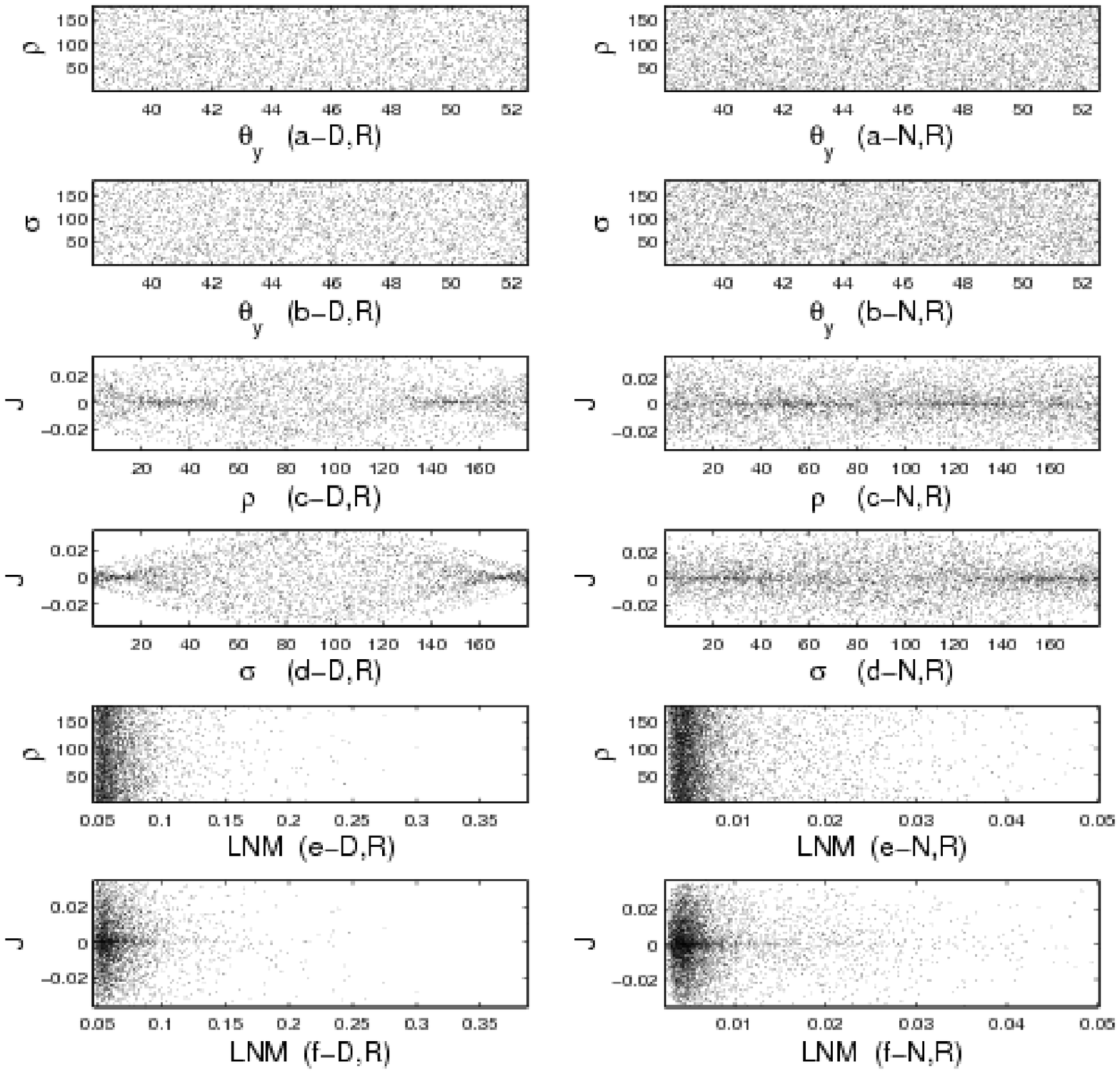}}
\end{minipage}
\vspace{0.5cm}
\caption{{\footnotesize Pattern $\mathbf C_{32}:$ Left panel presents correlations of $m_{ee}$ against
$\t_y$, $\d$,  $\r$ , $\s$, and $J$. It also shows the correlation between $m_2/m_3$ and $\t_y$.
 The right panel shows correlations of ($\r, \s$) against $\t_y$ and  $J$ and those of the
  lowest neutrino mass (LNM) versus $\r$ and $J$.}}
\label{c32fig2}
\end{figure}

\begin{figure}[hbtp]
\centering
\epsfxsize=10cm
\epsfbox{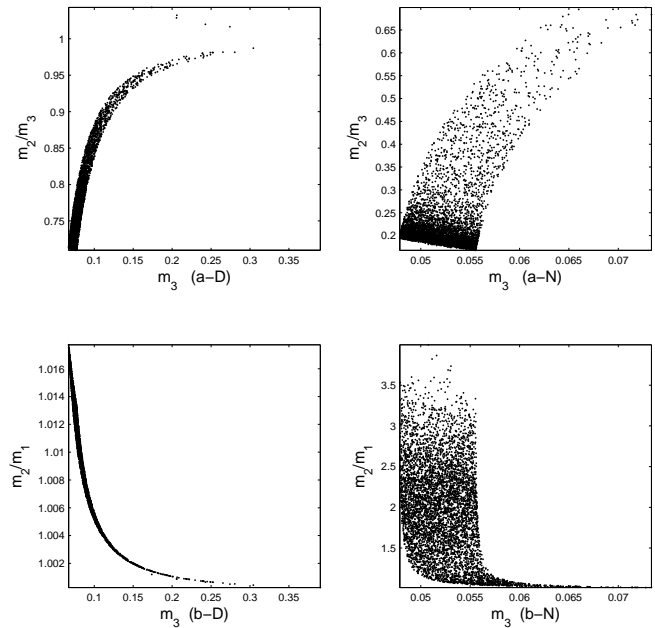}
\caption{{\footnotesize Pattern $\mathbf C_{32}:$ correlations of mass ratios ${m_2\over m_3}$ and
${m_2\over m_1}$ against $m_3$.}}
\label{c32fig3}
\end{figure}

Upon spanning the parameter space, we checked that no inverted hierarchy could accommodate the data. We
produce the correlation plots in Figures (\ref{c32fig1}, \ref{c32fig2}  and \ref{c32fig3}). We see that the
mixing angles and phase angles can cover their experimentally allowed regions. Linear correlations
between $\d$ and ($\r, \s$) are apparent in the degenerate case, whereas the linear correlation of $\r$
versus $\s$ is also apparent in the normal case. The `sinusoidal' and `isosceles' shapes of the
($J,\d$) and ($J,\t_z$) correlations are uniformly covered. Again, the correlations of $m_{ee}$ show that a
lower bound on this parameter restricts enormously the parameter space. These correlation-plots, or alternatively
Table~\ref{tab2}, show that the limit $m_{ee} = 0$ can be met in the normal hierarchy case. Moreover, the pattern
in this limit is a two-zero entries texture with $M_{\n11} = M_{\n12} = 0$ or $M_{\n11} = M_{\n13} = 0$. The equivalent
models for two vanishing minors texture \cite{LashinChamoun} are the $T_1$-symmetry related models: (vanishing $C_{33}$
\& $C_{32}$) and (vanishing $C_{22}$ \& $C_{32}$).
Again, no clear correlation between ($m_{23}, \t_y$), nor between $\t_y$ and ($\r, \s$). One can find a mild
correlation of $J$ versus $\r$ and $\s$, originating from the linear correlation of $\d$ with $(\r, \s)$, especially in the degenerate case. As to the LNM correlations, the trend is to favor a lower value for this parameter in that increasing its value would cut short the parameter space. Numerically, the lower bounds on $\t_z = 0$ reached very tiny values in this pattern (look at
Table~\ref{tab2}).

For the mass spectrum, the normal hierarchy is not acute, in that the ratio $m_2/m_3$ has a lower bound of order $0.2$
(plot a-N in Fig~\ref{c32fig3}). We note also that no mass can approach too closely to zero. We see this in the normal hierarchy
either by looking at (Fig~\ref{c32fig3}, plot b-N) and noting that $\frac{m_2}{m_1}$ is not reaching very large values
 corresponding to very minute $m_1$, or by checking the coverable mass regions in Table~\ref{tab2}.

There is no non-invertible such texture which can accommodate the current data.

\subsection{  Pattern of vanishing minor $\mathbf {C_{21};\; M_{\nu 21}\,M_{\nu 33}- M_{\nu 31}\,M_{\nu 23}=0}$}
The relevant expressions for $A_1$, $A_2$ and $A_3$ for this model
are
\bea
A_1 &=& c_z c_x (c_y s_x + s_z c_x s_y\, e^{-i\,\d})\,e^{-i\,\d}\nn \\
A_2 & = & -c_z s_x (c_y c_x - s_z s_x s_y\, e^{-i\,\d})\,e^{-i\,\d},\nn\\
A_3 &=& s_z c_z c_y e^{-2\,i\,\d},
\label{minc21}
\eea
We get
\bea
\frac{m_1}{m_3} &\approx& \frac{c_x s_x s_{2\s - 2\r}}{ s_{2\s - \d}\, s_z \,t_y} +
O \left( 1 \right) \\
\frac{m_2}{m_3} &\approx& \frac{c_x s_x s_{2\s - 2\r}}{s_{\r - \d}\,s_z\,t_y}
 +O \left( 1 \right)
 \eea

with \bea R_\nu &=& \frac{c_{2\r} c_{2\r -2\d}-c_{2\s} c_{2\s -2\d}}{c_{2\s}c_{2\s -2\d}-c_\d^2}+O \left(
s_z
 \right)
\eea

The phenomenological analysis of this pattern can be deduced from that of $C_{31}$ which is equivalent
under the symmetry $T_1$.

Also, and as in the pattern $C_{31}$, there is a non-invertible such texture which can accommodate the current data, and this
happens only when $\t_z=0$ leading to $m_3=0$.


\subsection{  Pattern of vanishing minor $\mathbf {C_{11};\; M_{\nu 22}\,M_{\nu 33}- M_{\nu 32}\,M_{\nu 23}=0}$}

The quantities $A_1$, $A_2$ and $A_3$, corresponding to the model, are
\bea
A_1=c_z^2 s_x^2\, e^{-2\,i\d}, & A_2 = c_z^2 s_x^2\, e^{-2\,i\d},
& A_3= s_z^2\, e^{-2\,i\d}.
\label{minc11}
\eea
The analytical expressions for all relevant computed parameters are simple and
independent of $\d$.
The mass ratios take the forms:
 \bea
\frac{m_1}{m_3} &=&  \frac{c_x^2 s_{2\r-2\s}}{t_z^2 s_{2\s}} \\
\frac{m_2}{m_3} &=& \frac{s_x^2 s_{2\s-2\r}}{t_z^2 s_{2\r}},
\label{mc11}
\eea
Fixing the $\D m_{\mbox{sol}}^2$ at its central value (eq.(\ref{Deltam})), one can compute
$m_3$:
\be
m_3 = {{\sqrt{\Delta m^2_{\mbox{sol}}}\,
t_z^2}\over {\left|s_{2\s-2\r}\right|\;\sqrt{\left|\left(\frac{s_x^4}{s_{2\r}^2}-
{c_x^4\over s_{2\s}^2}\right)\right|}}},
\label{m3}
\ee
We thus can get the corresponding expression  of
$\D m_{\mbox{atm}}^2$ as:
\be
\D m_{\mbox{atm}}^2 = m_3^2 \left|1-{s_x^4\, s_{2\s-2\r}^2 \over
t_z^4 s_{2\r}^2}\right|
\ee

The non oscillation parameters $\langle m\rangle_e $, $\langle m\rangle_{ee} $
and $\Sigma$ are given as
\bea
\langle m\rangle_e &=& m_3\;\sqrt{\left[ {c_z^2\over t_z^4}\;s_{2\s-2\r}^2
\left({c_x^6\over s_{2\s}^2}+ {s_x^6\over s_{2\r}^2}\right) +
s_z^2\right]},\nonumber\\
\langle m\rangle_{ee}&=& m_3\;\left| {c_x^2\over t_z^2}
{s_{2\r-2\s}\over s_{2\s}}\, c_x^2 c_z^2 e^{2\,i\r} + {s_x^2\over t_z^2}
{s_{2\s-2\r}\over s_{2\r}}\, s_x^2 c_z^2 e^{2\,i\s} +
s_z^2\right|, \label{meec11}\\
\Sigma &=& m_3\;\left |{c_x^2\over t_z^2}
{s_{2\r-2\s}\over s_{2\s}} + {s_x^2\over t_z^2}
{s_{2\s-2\r}\over s_{2\r}} + 1\right | \nonumber.
\eea
where $m_3$ is given in eq.(\ref{m3}). Finally the parameter $R_\nu$
has the form:
\be
R_\nu = {s^2_{2\s-2\r}\,\left({s_x^4\over s_{2\r}^2} - {c_x^4\over s_{2\s}^2}\right) \over
1 - {s_x^4\, s^2_{2\s-2\r} \over t_z^4\,s_{2\r}^2}}
\ee

\begin{figure}[hbtp]
\centering
\begin{minipage}[l]{0.5\textwidth}
\epsfxsize=8.8cm
\centerline{\epsfbox{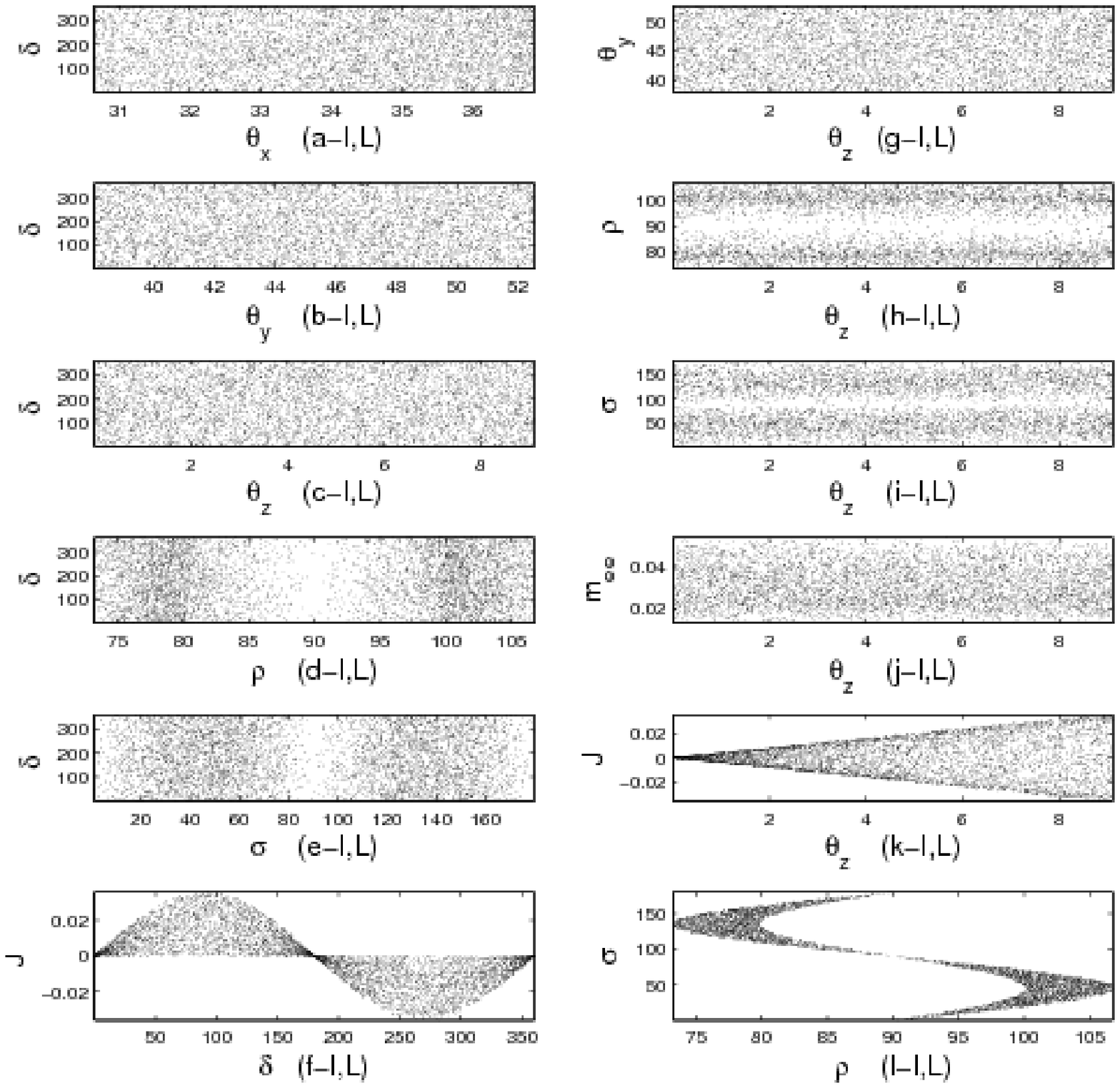}}
\end{minipage}%
\begin{minipage}[r]{0.5\textwidth}
\epsfxsize=8.8cm
\centerline{\epsfbox{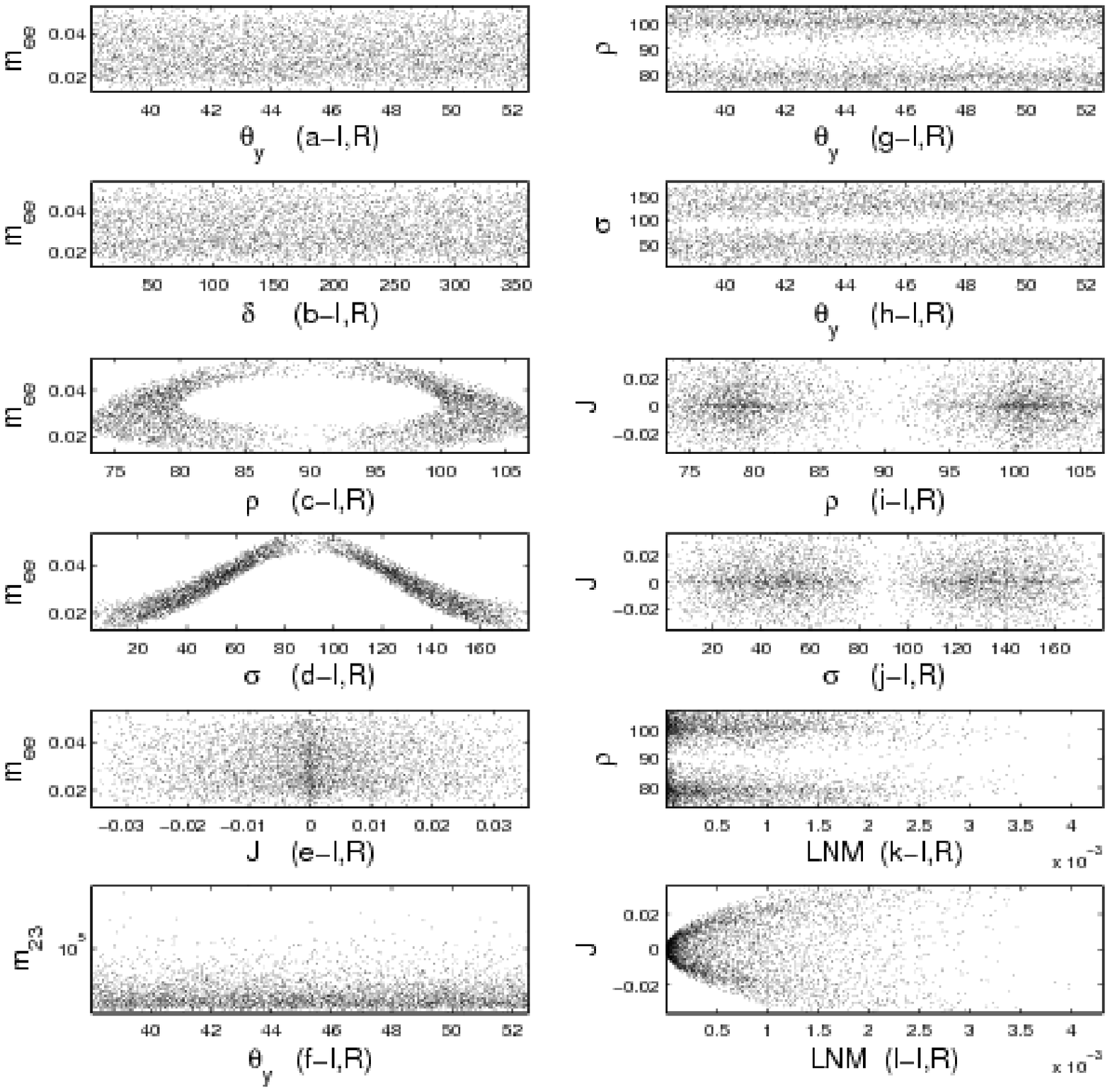}}
\end{minipage}
\vspace{0.5cm}
\caption{{\footnotesize Pattern $\mathbf C_{11}:$ Left panel presents, in the first column, correlations
of $\delta$ against
mixing angles, CP-phases and $J$. It also shows, in the second column, correlations of $\th_z$ against $\th_y$, $\r$ ,
$\s$, $m_{ee}$ and $J$, and the correlation of $\r$ versus $\s$. Right panel shows, in the third column,
correlations of $m_{ee}$ against
$\t_y$, $\d$,  $\r$ , $\s$, and $J$, and also the correlation between $m_2/m_3$ and $\t_y$.
It presents in the last column correlations of ($\r, \s$) against $\t_y$ and $J$ and those of the
  lowest neutrino mass (LNM) versus $\r$ and $J$.}}
\label{c11fig1}
\end{figure}

\begin{figure}[hbtp]
\centering
\epsfxsize=10cm
\epsfbox{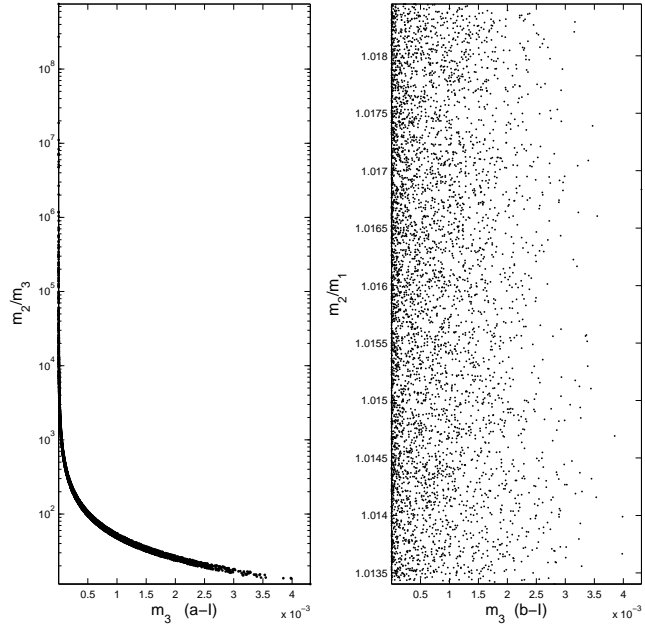}
\caption{{\footnotesize Pattern $\mathbf C_{11}:$ correlations of mass ratios ${m_2\over m_3}$ and
${m_2\over m_1}$ against $m_3$.}}
\label{c11fig3}
\end{figure}

This pattern shows only inverted-type hierarchy, and the corresponding plots are shown in
Figures (\ref{c11fig1} and \ref{c11fig3}). We see in Figure~\ref{c11fig1} that the mixing angles ($\t_x,\t_y,\t_z$)
and the Dirac phase angle $\d$
cover all their allowable regions (plots: a-L $\rightarrow$ c-L, g-L $\rightarrow$ i-L). However, the region
around $\r = \frac{\pi}{2}$ or $\s = \frac{\pi}{2}$ tends to be excluded (look also at the two plots: g-R, h-R), in
accordance with the analytic formulae, say (eq~\ref{mc11}) where these limits would equate the denominators to zero.
Moreover, the analytic expressions (e.g. eq~\ref{m3}) would exclude the region of $\r-\s$ equal to
a multiple of $\frac{\pi}{2}$. Furthermore, setting the ratio $m_2/m_1$ to be larger than $1$ and taking into
account that $t_x$ is less than one, for the experimentally accepted $\t_x$, would force the ratio
$|\frac{s_{2\s}}{s_{2\r}}|$ to be larger than one. This, with the fact that the difference $\r-\s$ should not vanish,
would put a lower bound on $\s$, as one can
see in Table~ \ref{tab2}. In this table we see also a restricted region for $\r$ due to the $\Delta_{\mbox{atm}}^2$ formula
barring small values of $\r$.
Plots (d-L, e-L) show no strong correlation between ($\d, \r$), nor between ($\d, \s$), whence no
clear correlation between $J$ versus $\r$, or between $J$ versus $\s$ (plots: i-R, j-R).
There is a strong `sinusoidal' correlation between $\s$ against $\r$ (plot: l-L) showing that $\s$ being in the
first quarter forces $\r$ to be in the second quarter, and vice versa. The correlations of $m_{ee}$ have a
clear shape only versus $\r$ and $\s$. These shapes can be deduced from the analytical formula
(equation \ref{meec11}). The LNM correlation with $\r$ (plot k-R) again excludes the region around $\r = \frac{\pi}{2}$,
whereas its correlation with $J$ favors, for very small values of LNM, a vanishing $J$ with no CP violation effects.

The mass spectrum in Figure~\ref{c11fig3} shows a quite strong inverted hierarchy (plot a-I) with
 $m_1 \sim m_2$ (plot b-I), and that we can approach the limit $m_3 = 0$. In fact, there is a viable singular such texture when $\t_z=0$ and $m_3=0$.

\begin{landscape}
\begin{table}[h]
 \begin{center}
{\tiny
 \begin{tabular}{c|c|c|c|c|c|c|c|c|c|c|c|c}
 \hline
\multicolumn{13}{c}{\mbox{Model} $C_{33}: M_{\nu 11}\,M_{\nu 22}- M_{\nu 12}\,M_{\nu 12}=0$} \\
\hline
 \mbox{quantity} & $\th_x$ & $\th_y$& $\th_z$ & $m_1$ & $m_2$& $m_3$ & $\r$ & $\sig$ & $\d$ & $\me$
 & $\mee$ & $J$\\
 \hline
 \multicolumn{13}{c}{\mbox{Degenerate  Hierarchy}} \\
 \cline{1-13}
 $1\, \sig$ &$31.94 - 34.45$ & $42.13 - 48.44$ & $0.0003 - 6.28$ & $0.0489 - 0.4029$ & $0.0497 - 0.4030$ &
  $0.0659 - 0.3997$& $0.0780 - 179.83$ & $0.1143 - 179.69$&$0.1978 - 359.83$ & $0.0493 - 0.4029$
  &$0.0222 - 0.3754$ & $-0.0251 - 0.0246$ \\
 \hline
 $2\, \sig$ & $30.65 - 36.86$ & $39.65 - 52.54$& $0.0004 - 9.1$ & $0.0471 - 0.4027$ & $0.0479 - 0.4028$ &
 $0.0567 -  0.3996$&$0.2870 - 179.88$ & $0.0715 - 179.93$ & $0.0554 - 359.90$& $0.0475 - 0.4027$ &
 $0.0165 - 0.3309$ & $-0.0348 - 0.0356$  \\
 \hline
 $3\, \sig$ &$29.34 - 39.23$ &$38.95 - 55.55$ & $0.0008 - 11.53$ & $0.0449 - 0.4046$ & $0.0458 - 0.4047$ &
  $0.0543 - 0.4072$& $0.0008 - 180$ & $0.1194 - 180$ & $0.3829 -  359.57$ &
  $0.0453 - 0.4047$ & $0.0097 - 0.2884$ & $-0.0459 - 0.0455$ \\
 \hline
 \multicolumn{13}{c}{\mbox{Normal  Hierarchy}} \\
 \cline{1-13}
 $1\, \sig$ &$31.94 - 34.45$& $42.13 - 48.44$ &$0.0000 - 6.29$ &$0.0019 -  0.0491$ &$0.0091 - 0.0499$ &
  $0.0499 - 0.0721$& $0.0056 -179.98$ &$0.0493 - 179.97$ & $0.0307 - 359.87$ &$0.0056 - 0.0494$ &
  $0.0000 -  0.0422$ & $-0.0249 - 0.0247$  \\
 \hline
 $2\, \sig$ & $30.65 - 36.87$ &$38.05 - 52.54$ & $0.0025 - 9.1$ &$0.0013 - 0.0510$ &$0.0090 -  0.0518$
 &$ 0.0478 - 0.0742$ & $0.1356 - 179.95$ & $ 0.0515 - 179.99$ & $0.1023 - 359.99$& $0.0052 - 0.0513$ &
 $0.0000 -   0.0467$ & $-0.0365 - 0.0359$  \\
 \hline
 $3\, \sig$ &$29.33 - 39.23$ & $35.66 - 55.55$ & $0.0022 - 11.53$& $0.0008 - 0.0514$ & $0.0089 - 0.0522$ &
  $0.0456 - 0.0756$& $0.0465 - 180$ & $0.0091 - 179.99$ &$0.1535 - 359.99$ & $0.0050  - 0.0523$ &
  $0.0000 - 0.0449$ & $-0.0453 - 0.0460$ \\
 \hline
 \multicolumn{13}{c}{\mbox{Inverted  Hierarchy}} \\
 \cline{1-13}
 $1\, \sig$ & $\times$  &$\times$ &$\times$ &$\times$ &$\times$ &$\times$
 &$\times$ &$\times$ &$\times$ &$\times$ &$\times$ &$\times$  \\
 \hline
 $2\, \sig$ &$30.65 - 36.87$ & $48.41 - 52.54$ & $0.0137 -  9.095$ & $0.0575 - 0.0839$ &
 $0.0582 - 0.0844$ & $0.0339 - 0.0645$ & $0.0734 - 179.88$ & $0.0687 -  179.61$ & $0.0338 - 359.66$ &
  $0.0573 - 0.0840$& $ 0.0444 -  0.0831$ & $-0.0357 -  0.0352$  \\
 \hline
 $3\, \sig$ & $29.33 - 39.23$ & $48.04 - 55.55$ & $0.0003 - 11.54$ & $0.0499 - 0.0865$ &
 $0.0506 - 0.0870$ & $0.0229  - 0.0662$ & $0.0213 - 179.97$ & $0.0095 - 179.96$ & $0.1354 - 359.83$ &
  $0.0495 - 0.0866$ & $0.0300 - 0.0853$& $ -0.0442 - 0.0442$  \\
 \hline
 \hline
\multicolumn{13}{c}{\mbox{Model} $C_{22}: M_{\nu 11}\,M_{\nu 33}- M_{\nu 13}\,M_{\nu 31}=0$} \\
\hline
 \mbox{quantity} & $\th_x$ & $\th_y$& $\th_z$ & $m_1$ & $m_2$& $m_3$ & $\r$ & $\sig$ & $\d$ & $\me$
 & $\mee$ & $J$\\
 \hline
 \multicolumn{13}{c}{\mbox{Degenerate  Hierarchy}} \\
 \cline{1-13}
 $1\, \sig$ &$31.94 - 34.45$ & $42.13 - 48.44$ & $0.0023 - 6.2858$ & $0.0489 - 0.4563$ & $0.0497 - 0.4563$ &
  $0.0699 - 0.4593$& $0.0784 - 179.98$ & $0.0779 - 179.91$&$0.2430 - 359.93$ & $0.0494 - 0.4563$
  &$0.0223  - 0.3911$ & $-0.0248 - 0.0247$ \\
 \hline
 $2\, \sig$ & $30.65 - 36.87$ & $38.05 - 50.27$& $0.0013  - 9.1$ & $0.0469 - 0.4206$ & $0.0477 - 0.4207$ &
 $0.0575 - 0.4178$&$0.0508 - 179.98$ & $0.0325 - 179.90$ & $0.0793 - 359.81$& $0.0473 - 0.4206$ &
 $0.0159 - 0.3269$ & $-0.0359  - 0.0361$  \\
 \hline
 $3\, \sig$ &$29.33 - 39.23$ &$35.66 -  50.96$ & $0.0007 - 11.53$ & $0.0451 - 0.4527$ & $0.0459 - 0.4528$ &
  $0.0552 - 0.4559$& $0.0273 - 179.93$ & $0.0474 - 179.95$ & $0.2755 - 359.85$ &
  $ 0.0454 - 0.4527$ & $ 0.0109 - 0.3794$ & $-0.0450 - 0.0456$ \\
 \hline
 \multicolumn{13}{c}{\mbox{Normal  Hierarchy}} \\
 \cline{1-13}
 $1\, \sig$ &$31.94 - 34.45$& $42.13 - 48.44$ &$0.0024 - 6.28$ &$0.0017 - 0.0487$ &$0.0091 - 0.0495$ &
  $0.0499 - 0.0720$& $0.1438 - 179.97$ &$0.0954 - 179.95$ & $0.0441 - 359.73$ &$ 0.0055 - 0.0491$ &
  $0.0000  -  0.0397$ & $-0.0248 - 0.0242$  \\
 \hline
 $2\, \sig$ & $30.65 - 36.87$ &$38.06 -  52.54$ & $0.0004 - 9.1$ &$0.0013 - 0.0519$ &$0.0090 - 0.0527$
 &$ 0.0478 - 0.0759$ & $0.0287 - 179.97$ & $ 0.0578 - 179.97$ & $0.0066 - 359.99$& $0.0052 - 0.0523$ &
 $0.0000 - 0.0374$ & $-0.0355  -  0.0359$  \\
 \hline
 $3\, \sig$ &$29.33 - 39.23$ & $35.66 - 55.55$ & $0.0002 - 11.53$& $0.0006 - 0.0514$ & $0.0089 - 0.0521$ &
  $0.0456 - 0.0748$& $0.0168 - 179.97$ & $0.0071 - 180$ &$0.0304 - 359.98$ & $ 0.0050 -  0.0525$ &
  $0.0000 - 0.0454$ & $-0.0457 - 0.0451$ \\
 \hline
 \multicolumn{13}{c}{\mbox{Inverted  Hierarchy}} \\
 \cline{1-13}
 $1\, \sig$ & $\times$  &$\times$ &$\times$ &$\times$ &$\times$ &$\times$
 &$\times$ &$\times$ &$\times$ &$\times$ &$\times$ &$\times$  \\
 \hline
 $2\, \sig$ &$30.65 - 36.87$ & $38.05 - 41.50$ & $0.0006 - 9.1$ & $0.0582 - 0.0842$ &
 $0.0589 - 0.0847$ & $0.0352 - 0.0648$ & $0.0156 - 179.89$ & $0.0305 - 179.93$ & $0.0713 - 359.80$ &
  $0.0580 - 0.0842$& $ 0.0457 - 0.0826$ & $-0.0353 - 0.0348$  \\
 \hline
 $3\, \sig$ & $29.33 - 39.23$ & $35.66 - 41.77$ & $0.0021 - 11.53$ & $0.0514 - 0.0865$ &
 $0.0522  - 0.0869$ & $0.0265  - 0.0664$ & $0.0254 - 179.96$ & $0.0049 - 179.87$ & $0.1171 - 360$ &
  $0.0510 - 0.0865$ & $0.0320 - 0.0854$& $-0.0436  -  0.0447$  \\
 \hline
 \hline
\multicolumn{13}{c}{\mbox{Model} $C_{31}: M_{\nu 12}\,M_{\nu 23}- M_{\nu 13}\,M_{\nu 22}=0$} \\
\hline
 \mbox{quantity} & $\th_x$ & $\th_y$& $\th_z$ & $m_1$ & $m_2$& $m_3$ & $\r$ & $\sig$ & $\d$ & $\me$
 & $\mee$ & $J$\\
 \hline
 \multicolumn{13}{c}{\mbox{Degenerate  Hierarchy}} \\
 \cline{1-13}
 $1\, \sig$ &$31.94 -  34.45$ & $42.14 - 48.44$ & $ 0.0542 - 6.29$ & $0.0493 - 0.2988$ &$0.0501 - 0.2989$ &
  $0.0593  - 0.2946$& $0.0050 - 179.98$ & $0.0040 - 1780$ &$0.3220 - 359.93$ & $0.0495 - 0.2989$
  &$0.0489 - 0.2988$ & $-0.0246 -  0.0245$ \\
 \hline
 $2\, \sig$ & $30.65 - 36.87$ & $38.055 - 52.53$& $0.0715 -  9.01$ & $0.0473 - 0.4431$ &$0.0481 - 0.4432$ &
 $0.0569 - 0.4399$&$ 0.0017 - 179.99$ & $ 0.0161 - 179.97$ & $0.0778 - 359.83$& $0.0476 - 0.4431$ &
 $0.0458 - 0.4368$ & $-0.0352 - 0.0360$  \\
 \hline
 $3\, \sig$ &$29.33  - 39.23$ & $35.66 - 55.54$ & $0.0705 - 11.54$& $0.0450 - 0.4217$ & $0.0459 - 0.4217$ &
  $0.0543 - 0.4184$& $0.0200 - 179.99$ & $ 0.0040 -  179.99$ &$0.0673 - 359.84$ & $0.0455 - 0.4216$ &
  $0.0395 - 0.4216$ & $-0.0442 - 0.0453$ \\
 \hline
 \multicolumn{13}{c}{\mbox{Normal  Hierarchy}} \\
 \cline{1-13}
 $1\, \sig$ &$31.94 - 34.45$& $42.13 - 48.44$ &$0.3175 - 6.29$ &$0.0092 - 0.0504$ &$0.0128 - 0.0511$ &
  $0.0509  - 0.0732$& $0.0229 - 179.96$ &$0.0043 - 179.94$ & $0-123\cup 242-360$ &$0.0117 - 0.0507$ &
  $0.0096 - 0.0506$ & $-0.0249 - 0.0251$  \\
 \hline
 $2\, \sig$ & $30.65 -  36.87$ &$38.06  - 52.54$ & $0.3599 -  9.09$ &$0.0061 - 0.0521$ &$0.0108 - 0.0529$
 &$ 0.0483 - 0.0760$ & $0.0486 - 179.97$ & $0.0251 - 179.85$ & $0-123\cup 242-360$& $ 0.0105 - 0.0528$ &
 $0.0062 - 0.0524$ & $-0.0356 - 0.0363$  \\
 \hline
 $3\, \sig$ &$29.33 - 39.23$ & $35.67 - 55.55$ & $0.2618 - 11.54$& $0.0047 - 0.0543$ & $0.0101 - 0.0550$ &
  $0.0462 - 0.0787$& $0.0018 - 179.98$ & $ 0.0258 -  179.99$ &$0-123\cup 242-360$ & $0.0101 - 0.0551$ &
  $0.0047 - 0.0538$ & $-0.0450 - 0.0456$ \\
 \hline
 \multicolumn{13}{c}{\mbox{Inverted  Hierarchy}} \\
 \cline{1-13}
 $1\, \sig$ &$31.94 - 34.45$ & $42.13 - 48.44$ & $0.0009  -  6.29$ & $0.0482  -  0.0804$ &
 $0.0490 - 0.0809$ & $ 0.0000 -  0.0615$ & $ 0.0428 - 180$ & $0.0583 - 179.97$ & $0.0081 - 359.89$ &
  $0.0484 - 0.0804$& $ 0.0183  -   0.0789$ & $-0.0246 -   0.0249$  \\
 \hline
 $2\, \sig$ &$30.65 - 36.87$ & $ 38.06 - 52.54$ & $0.0000 - 9.01$ & $0.0461 - 0.0809$ &
 $0.0469 -  0.0814$ & $ 0.0000 -  0.0620$ & $ 0.0224 - 179.95$ & $0.0362 - 179.95$ & $0.0358 - 359.97$ &
  $0.0463 - 0.0809$& $ 0.0141 - 0.0806$ & $-0.0357 - 0.0360$  \\
 \hline
 $3\, \sig$ & $29.33 - 39.23$ & $35.66 - 55.55$ & $0.0009 - 11.54$ & $0.0439 - 0.0858$ &
 $0.0447 - 0.0862$ & $0.0000 -  0.0654$ & $0.0020 - 179.99$ & $ 0.0127 - 179.99$ & $0.0026 - 359.94$ &
  $0.0441 - 0.0857$ & $0.0101 - 0.0848$& $-0.0459 - 0.0456$  \\
 \hline

 \end{tabular}
 }
 \end{center}
  \caption{\small \label{tab1} The various prediction for the models of
  one-vanishing minor $C_{33}$ , $C_{22}$ and $C_{31}$. The minor corresponding to the index $(ij)$
  is the determinant of the sub-matrix obtained by deleting the
  $i^{th}$ line and the $j^{th}$ column. All the angles (masses) are
  evaluated in degrees ($eV$). The mark $\times$ indicates that the corresponding pattern with the specified hierarchy type can not accommodate the experimental data at the given $\s$ precision level.}
 \label{modelc33}
  \end{table}
\end{landscape}

\begin{landscape}
\begin{table}[h]
 \begin{center}
{\tiny
 \begin{tabular}{c|c|c|c|c|c|c|c|c|c|c|c|c}
 \hline
\multicolumn{13}{c}{\mbox{Model} $C_{32}: M_{\nu 11}\,M_{\nu 23}- M_{\nu 21}\,M_{\nu 13}=0$} \\
\hline
 \mbox{quantity} & $\th_x$ & $\th_y$& $\th_z$ & $m_1$ & $m_2$& $m_3$ & $\r$ & $\sig$ & $\d$ & $\me$
 & $\mee$ & $J$\\
 \hline
 \multicolumn{13}{c}{\mbox{Degenerate  Hierarchy}} \\
 \cline{1-13}
 $1\, \sig$ &$31.94 - 34.45$ & $42.13 - 48.44$ & $0.0007 - 6.29$ & $0.0489 - 0.3459$& $0.0497 - 0.3460$ &
  $0.0699  -  0.3495$ & $0.0296 - 179.94$&$ 0.1237 - 180$ & $0.0125 - 359.98$
  &$0.0493 - 0.3459$ & $ 0.0336 - 0.3419$ & $-0.0250 - 0.0248$\\
 \hline
 $2\, \sig$ & $30.66 - 36.87$ & $38.05 - 52.54$& $0.0029 - 9.09$ & $0.0470 - 0.3870$ & $0.0478 - 0.3871$ &
 $0.0670 - 0.3902$& $0.0101 - 179.83$ & $0.0444 - 179.92$ & $0.1492 - 359.99$& $0.0475 - 0.3870$ &
 $0.0293 - 0.3810$ & $-0.0362 - 0.0348$  \\
 \hline
 $3\, \sig$ &$29.34 - 39.23$ &$35.67 -  55.54$ & $0.0006 - 11.53$ & $0.0449 - 0.3765$ & $0.0458 - 0.3766$ &
  $0.0641 - 0.3738$ & $0.0774 - 179.87$ & $0.0061 - 180$ & $0.0510 - 359.93$ &
  $0.0453 - 0.3764$ & $0.0232 -  0.3396$ & $-0.0453 - 0.0459$ \\
 \hline
 \multicolumn{13}{c}{\mbox{Normal  Hierarchy}} \\
 \cline{1-13}
 $1\, \sig$ &$31.94 - 34.45$& $42.13 - 48.44$ &$0.0001 - 6.29$ &$0.0029 - 0.0504$ &$0.0093 - 0.0512$ &
  $0.0499  - 0.0733$& $0.0083 - 179.92$ &$0.0003 - 179.97$ & $0.1523 - 359.99$ &$0.0056 - 0.0508$ &
  $0.0001 - 0.0352$ & $-0.0245 - 0.0251$  \\
 \hline
 $2\, \sig$ & $30.65 - 36.87$ &$38.05 - 52.54$ & $0.0019 - 9.1$ &$0.0023 - 0.0501$ &$ 0.0092 - 0.0509$
 &$ 0.0479 -  0.0734$ & $0.0047 - 179.97$ & $ 0.0596 - 179.99$ & $0.0083 - 359.97$& $0.0053 - 0.0507$ &
 $0.0000 - 0.0346$ & $-0.0356 - 0.0366$  \\
 \hline
 $3\, \sig$ &$29.33 - 39.23$ & $35.66 - 55.55$ & $0.0021 - 11.54$& $0.0019 - 0.0525$ & $0.0091 - 0.0532$ &
  $0.0457 -  0.0776$& $0.0185 - 179.99$ & $0.0077 - 179.96$ &$0.0224 - 360$ & $0.0050  - 0.0530$ &
  $0.0000 - 0.0365$ & $-0.0455 - 0.0444$ \\
 \hline
 \multicolumn{13}{c}{\mbox{Inverted  Hierarchy}} \\
 \cline{1-13}
 $1\, \sig$ & $\times$  &$\times$ &$\times$ &$\times$ &$\times$ &$\times$
 &$\times$ &$\times$ &$\times$ &$\times$ &$\times$ &$\times$  \\
 \hline
 $2\, \sig$ & $\times$  &$\times$ &$\times$ &$\times$ &$\times$ &$\times$
 &$\times$ &$\times$ &$\times$ &$\times$ &$\times$ &$\times$  \\
 \hline
 $3\, \sig$ & $\times$  &$\times$ &$\times$ &$\times$ &$\times$ &$\times$
 &$\times$ &$\times$ &$\times$ &$\times$ &$\times$ &$\times$  \\
 \hline
 \hline
\multicolumn{13}{c}{\mbox{Model} $C_{21}: M_{\nu 21}\,M_{\nu 33}- M_{\nu 31}\,M_{\nu 23}=0$} \\
\hline
 \mbox{quantity} & $\th_x$ & $\th_y$& $\th_z$ & $m_1$ & $m_2$& $m_3$ & $\r$ & $\sig$ & $\d$ & $\me$
 & $\mee$ & $J$\\
 \hline
 \multicolumn{13}{c}{\mbox{Degenerate  Hierarchy}} \\
 \cline{1-13}
 $1\, \sig$ &$31.94 - 34.45$ & $42.13 - 48.43$ & $ 0.0386 - 6.29$ & $0.0492 -0.4521$ & $0.0500 - 0.4521$ &
  $0.0593 - 0.4549$& $0.0026 - 180$ & $0.0098 -  179.99 $&$0.1274 - 359.11$ & $0.0494 - 0.4521$
  &$0.0487 - 0.4521$ & $-0.0245 - 0.0248$ \\
 \hline
 $2\, \sig$ & $30.65 - 36.87$ & $ 38.05 - 52.54$& $0.0524 - 9.1$ & $0.0470 - 0.4365$ & $0.0478 - 0.4366$ &
 $0.0568 -  0.4335$& $0.0014 - 180$ & $0.0018 - 179.98$ & $0.2779 -  359.63$& $0.0473 - 0.4365$ &
 $0.0456 -  0.4365$ & $ -0.0362 - 0.0361$  \\
 \hline
 $3\, \sig$ &$29.33 - 39.23$ &$35.68 - 55.55$ & $0.0185 - 11.53$ & $0.0451 - 0.3307$ & $0.0460 - 0.3308$ &
 $0.0542 -  0.3272$& $0.0118 - 180$ & $0.0003 -  179.98$ & $0.5547 - 359.84$ &
 $ 0.0456 - 0.3306$ & $ 0.0402 -  0.3306$ & $-0.0455 - 0.0455$ \\
 \hline
 \multicolumn{13}{c}{\mbox{Normal  Hierarchy}} \\
 \cline{1-13}
 $1\, \sig$ &$31.94 - 34.45$& $ 42.13 -  48.44$ &$0.2922 -  6.29$ &$0.0096 -  0.0505$ &$0.0131 - 0.0513$ &
  $0.0508 - 0.0734$& $ 0.0543 - 179.99$ &$0.0623 - 179.98$ & $59.58 - 304.26$ &$ 0.0120 - 0.0509$ &
  $0.0099 -  0.0506$ & $-0.0249 -  0.0248$  \\
 \hline
 $2\, \sig$ & $ 30.65 - 36.87$ &$38.06 - 52.54$ & $0.3368 - 9.1$ &$0.0062 - 0.0519$ &$0.0108 - 0.0526$
 &$  0.0483 -  0.0753$ & $0.0083 - 179.99$ & $ 0.1329 - 179.99$ & $57.51 - 296.59$& $0.0108 - 0.0527$ &
 $0.0066 - 0.0525$ & $-0.0363 - 0.0359$  \\
 \hline
 $3\, \sig$ &$29.33 - 39.23$ & $35.66 - 55.55$ & $0.4979 - 11.54$& $0.0050 -0.0527$ & $0.0102 - 0.0535$ &
  $0.0460 - 0.0778$& $0.0292 - 179.92$ & $0.0199 -  180$ &$59.53 - 302.99$ & $ 0.0099 - 0.0533$ &
  $0.0048 - 0.0530$ & $-0.0452 - 0.0457$ \\
 \hline
 \multicolumn{13}{c}{\mbox{Inverted  Hierarchy}} \\
 \cline{1-13}
 $1\, \sig$ & $31.94 - 34.45$  &$42.13 - 48.44$ &$0.0000 -  6.29$ &$0.0482 - 0.0806$ & $0.0490 - 0.0810$ &
 $0.0000  - 0.0619$ &$0.0079 - 179.95$ &$0.0857 - 179.96$ &$0.1136  - 359.99$ &$0.0484 - 0.0807$ &
 $ 0.0180 - 0.0804$ &$-0.0243 - 0.0246$  \\
 \hline
 $2\, \sig$ &$30.65 -  36.87$ & $38.05 - 52.54$ & $0.0026 -  9.09$ & $0.0461 - 0.0825$ &
 $0.0469 - 0.0830$ & $ 0.0000 - 0.0626$ & $0.0076 - 179.96$ & $0.0068 - 179.99$ & $0.0672 - 359.94$ &
  $0.0463 - 0.0824$& $ 0.0143 - 0.0811$ & $-0.0356 - 0.0352$  \\
 \hline
 $3\, \sig$ & $29.33 - 39.23$ & $35.66 - 55.55$ & $0.0012 - 11.52$ & $ 0.0438 - 0.0849$ &
 $0.0447 -  0.0853$ & $ 0.0000 -  0.0650$ & $0.0115 - 179.97$ & $0.0079 - 179.96$ & $0.3136 - 359.94$ &
  $0.0440 - 0.0849$ & $ 0.0094  - 0.0849$& $-0.0457 - 0.0443$  \\
 \hline
 \hline
\multicolumn{13}{c}{\mbox{Model} $C_{11}: M_{\nu 22}\,M_{\nu 33}- M_{\nu 32}\,M_{\nu 23}=0$} \\
\hline
 \mbox{quantity} & $\th_x$ & $\th_y$& $\th_z$ & $m_1$ & $m_2$& $m_3$ & $\r$ & $\sig$ & $\d$ & $\me$
 & $\mee$ & $J$\\
 \hline
 \multicolumn{13}{c}{\mbox{Degenerate  Hierarchy}} \\
 \cline{1-13}
 $1\, \sig$ & $\times$  &$\times$ &$\times$ &$\times$ &$\times$ &$\times$
 &$\times$ &$\times$ &$\times$ &$\times$ &$\times$ &$\times$  \\
 \hline
 $2\, \sig$ & $\times$  &$\times$ &$\times$ &$\times$ &$\times$ &$\times$
 &$\times$ &$\times$ &$\times$ &$\times$ &$\times$ &$\times$  \\
 \hline
 $3\, \sig$ & $\times$  &$\times$ &$\times$ &$\times$ &$\times$ &$\times$
 &$\times$ &$\times$ &$\times$ &$\times$ &$\times$ &$\times$  \\
 \hline
 \multicolumn{13}{c}{\mbox{Normal  Hierarchy}} \\
 \cline{1-13}
 $1\, \sig$ & $\times$  &$\times$ &$\times$ &$\times$ &$\times$ &$\times$
 &$\times$ &$\times$ &$\times$ &$\times$ &$\times$ &$\times$  \\
 \hline
 $2\, \sig$ & $\times$  &$\times$ &$\times$ &$\times$ &$\times$ &$\times$
 &$\times$ &$\times$ &$\times$ &$\times$ &$\times$ &$\times$  \\
 \hline
 $3\, \sig$ & $\times$  &$\times$ &$\times$ &$\times$ &$\times$ &$\times$
 &$\times$ &$\times$ &$\times$ &$\times$ &$\times$ &$\times$  \\
 \hline
 \multicolumn{13}{c}{\mbox{Inverted  Hierarchy}} \\
 \cline{1-13}
 $1\, \sig$ &$31.94 - 34.45$ & $42.13 - 48.44$ & $0.0006 - 6.29$ & $0.0482 - 0.0522$ &
 $0.0490 - 0.0529$ & $ 0.0000  - 0.0015$ & $ 76.21 - 103.81$ & $0.6212 - 178.88$ & $0.0156 - 359.88$ &
  $0.0482 - 0.0524$& $ 0.0177  - 0.0519$ & $-0.0246 - 0.0248$  \\
 \hline
 $2\, \sig$ &$ 30.65 - 36.87$ & $ 38.05 -  52.54$ & $0.0019 -  9.1$ & $0.0461 - 0.0541$ &
 $0.0469  - 0.0548$ & $  0.0000 - 0.0043$ & $ 73.23 - 106.74$ & $1.1273 - 179.71$ & $0.0851 - 359.9438$ &
  $0.0458 - 0.0543$& $ 0.0128 - 0.0534$ & $-0.0357 - 0.0358$  \\
 \hline
 $3\, \sig$ & $29.33 - 39.23$ & $35.66 -  55.54$ & $0.0026 -  11.53$ & $0.0438 - 0.0562$ &
 $0.0447 - 0.0569$ & $0.0000 -  0.0088$ & $69.51 -  110.52$ & $ 0.6325 - 178.33$ & $0.0559 - 359.89$ &
  $0.0434 - 0.0560$ & $0.0093 - 0.0543$& $-0.0452 - 0.0465$  \\
 \hline

 \end{tabular}
 }
 \end{center}
  \caption{\small \label{tab2} The various prediction for the models of
  one-vanishing minor $C_{32}$ , $C_{21}$ and $C_{11}$. The minor corresponding to the index $(ij)$
  is the determinant of the sub-matrix obtained by deleting the
  $i^{th}$ line and the $j^{th}$ column. All the angles (masses) are
  evaluated in degrees ($eV$). The mark $\times$ indicates that the corresponding pattern with the specified hierarchy type can not accommodate the experimental data at the given $\s$ precision level.}
  \label{modelc32}
 \end{table}
\end{landscape}

From the two tables we see that the value $m_3 = 0$ is attained in the inverted hierarchy for the patterns ($C_{11}$) and ($C_{31}$,$C_{21}$)
, the latter two being related by $T_1$-symmetry.

\section{Singular models}
The viable singular models obtained in cases $\mathbf C_{31}\equiv C_{21}$ and
and $\mathbf C_{11}$ are found only for vanishing $m_3$, albeit with zero
$\theta_z$. A vanishing $\theta_z$ is still consistent with
experimental data as shown in eq.~(\ref{osdata}). These models can accommodate the experimental data for the mixing
angles for any choice of the
 phase angles. It is interesting to notice
that the models $\mathbf C_{31}\equiv C_{21}$ and $\mathbf C_{11}$ have quite distinct
mass spectra but in the singular limit, $(m_3=0\; \mbox{and}\; \t_z=0)$,
the models become exactly identical leaving no room for any kind of distinguishability.

In fact, this model is identical to the singular model studied in \cite{LashinChamoun}
for the vanishing two-minors textures. For the sake of completeness we restate here the
 expressions for the mass parameters
\bea
m_1=\sqrt{\Delta m^2_{\mbox{atm}}- \Delta m^2_{\mbox{sol}}}, &m_2=\sqrt{\Delta m^2_{\mbox{atm}}},&
\langle m \rangle_{e}= \sqrt{m_1^2\,c_x^2 + m_2^2\,s_x^2} \\
\langle m \rangle_{ee}&=& \sqrt{\left| m_1^2\,c_x^4 + m_2^2\,s_x^4 +
2\,m_1\,m_2\, c_x^2\,s_x^2\,c_{2\rho-2\sigma}\right|},
\eea
and the mass matrix elements:

\bea
M_{\nu\;11} &= & \left( m_1\,c_x^2\,e^{2\,i\,\rho} +
m_2\,s_x^2\,e^{2\,i\,\sigma}\right),
\nonumber \\
M_{\nu\;12} &= & s_x\,c_x\,c_y\,e^{-i\,\d}\,
\left(-m_1\,e^{2\,i\,\rho} + m_2\,e^{2\,i\,\sigma}\right),\nonumber \\
M_{\nu\;13} &= & s_x\,c_x\,s_y\,e^{-i\,\d}\,
\left(m_1\,e^{2\,i\,\rho} - m_2\,e^{2\,i\,\sigma}\right),\nonumber \\
M_{\nu\;22} &= & c_y^2\,e^{-2\,i\,\d}\,
\left(m_1\,s_x^2\,e^{2\,i\,\rho} + m_2\,c_x^2\,e^{2\,i\,\sigma}\right),\nonumber \\
M_{\nu\;23} &= & -  c_y\,s_y\,e^{-2\,i\,\d}\,
\left(m_1\,s_x^2\,e^{2\,i\,\rho} + m_2\,c_x^2\,e^{2\,i\,\sigma}\right),\nonumber \\
M_{\nu\;33} &= & s_y^2\,e^{-2\,i\,\d}\,
\left(m_1\,s_x^2\,e^{2\,i\,\rho} + m_2\,c_x^2\,e^{2\,i\,\sigma}\right).
\label{m3zeroinv2}
\eea



\section{Symmetry realization}
All textures with  one zero-minor can be realized in a simple way in models based on seesaw mechanism
with a flavor Abelian symmetry. As mentioned earlier, if the Dirac neutrino mass matrix $M_D$ is diagonal
then a zero in the right-handed Majorana mass matrix $M_R$  leads to a zero minor in the effective
neutrino mass matrix $M_\n$.

We need three right-handed neutrinos $\nu_{Rj}$, three right-handed charged
leptons $l_{Rj}$ and three left-handed lepton doublets $D_{Lj}=(\nu_{Lj} , l_{Lj})^{T}$, where $j$
is the family index. Also we need the standard model (SM) Higgs, plus other scalar singlets.
We follow ~\cite{Lavoura} and assume a $Z_8$ underlying symmetry. For the sake of illustration,
let us take the case of $C_{33}$. Under the action of
$Z_8$, the leptons (the right singlets and the components of the left doublets) of the first, second and third families are multiplied by
$(1,-1,\om =\exp{({i\,\pi\over 4})})$ respectively, while the SM Higgs remains invariant.
This generates diagonal Dirac mass matrices for both charged leptons
and neutrinos.

The bilinears $\nu_{Ri} \nu_{Rj}$, relevant for the Majorana neutrino mass matrix
$M_R$, transform under $Z_8$  as
\be
\left(
\begin {array}{ccc}
1&\omega^4& \omega\\
\omega^4& 1& \omega^5\\
\omega&\omega^5&\omega^2
\end {array}
\right),
\label{Det1bi}
\ee
The $(1,1)$ and $(2,2)$ matrix elements of $M_R$ are $Z_8$
invariant, hence their corresponding mass terms are directly present in the
Lagrangian. We require a Yukawa coupling to a
real scalar singlet ($\chi_{12}$) which changes sign under $Z_8$ to generate the
$(1,2)$ matrix element in $M_R$, when acquiring a vev at the seesaw scale. The $(2,3)$
matrix element is equally generated by the Yukawa coupling to a complex
scalar singlet ($\chi_{23}$) with a multiplicative number $\omega^3$ under $Z_8$,
while the $(1,3)$ matrix element requires a Yukawa coupling to a complex
scalar singlet ($\chi_{13}$) which gets multiplied by $\omega^7$ under $Z_8$ .
The resulting right-handed Majorna mass matrix can be casted in the form,
\be
M_R=\left(
\begin {array}{ccc}
\times &\times& \times\\
\times& \times& \times\\
\times&\times&0
\end {array}
\right),
\label{Det1MR}
\ee
which is of the required form.

For the other patterns, they can be generated in a similar way summarized in the
table (\ref{Dettab3})
\begin{table}[hbtp]
\begin{center}
\begin{tabular}{||c||c|c|c|c|c|c|c|c|c|}
\hline
   \mbox{Model} & $1_F $ &$2_F $   & $3_F $ & $\chi_{11}$  &
   $\chi_{12}$& $\chi_{13}$ & $\chi_{22}$ & $\chi_{23}$ & $\chi_{33}$  \\
\hline
 $C_{33}$ & $1$ & $-1$ &  $\omega$ & absent   &
$-1$ & $\omega^7$ & absent & $\omega^2$ & absent
\\
\hline
  $C_{22}$ & $1$ & $\omega$ &  $-1$ & absent   &
 $\omega^7$ & $-1$ & absent & $\omega^3$ & absent\\
 \hline
 $C_{13}$ & $1$ & $\omega$ &  $-1$ & absent   &
 $\omega^7$ & absent & $\omega^6$ & $\omega^3$ & absent\\
\hline
 $C_{32}$ & $1$ & $\omega$ &  $-1$ & absent   &
 $\omega^7$ & $-1$ & $\omega^6$ & absent & absent\\
\hline
 $C_{12}$ & $1$ & $\omega$ &  $-1$ & absent   &
 absent & $-1$ & $\omega^6$ & $\omega^3$ & absent\\
\hline
 $C_{11}$ & $\omega$ & $-1$ &  $1$ & absent   &
 $\omega^3$ & $\omega^7$ & absent & $-1$ & absent\\
\hline
\end{tabular}
\end{center}
 \caption{\small  The $Z_8$symmetry realization for $6$ patterns of
 single vanishing minors. The index $1_F$ indicates the lepton first
 family and so on. The $\chi_{kj}$ denotes a scalar singlet which produces
 the entry $(k,j)$ of the right-handed Majorana mass matrix when acquiring vacuum
 expectation at the see-saw scale. The transformation properties, under the
 specified group, is listed below each family and needed scalar singlet for each
 model. $\omega$ denotes $\exp{({i\,\pi\over
4})}$, while $i=\sqrt{-1}$. }
\label{Dettab3}
 \end{table}

The models $C_{31}$, $C_{21}$ and $C_{11}$ in the limit of vanishing
$m_3$ and $\theta_z$ together give a singular mass matrix $M_\nu$. In fact, one
symmetry realization of such a singular model was stated in \cite{LashinChamoun}.

\section{ Discussion and conclusions}
We studied all the possible patterns of Majorana neutrino mass
matrices with one vanishing minor. All the six possible cases allow to accommodate the
current data without need to tune the input parameters. None of the patterns appears
as a `one-zero' texture and, for the chosen acceptable parameter points, all the
matrices are complex displaying CP violation effects.

Non-invertible mass matrices with one vanishing minor occur only in the cases
($C_{31}$, $C_{21}$ and $C_{11}$) with $\t_z=0$ leading to $m_3 =0$. These latter three cases coincide
in the limit $(\t_z \rightarrow 0)$  with the model of two vanishing minors in a
singular $M_\n$ which was studied in \cite{LashinChamoun}, making the distinction
between them difficult.

The model $C_{32}$ can not produce inverted-type hierarchy, whereas the model $C_{11}$ has only this type
of hierarchy. For all  six patterns, the mixing and phase angles cover their experimentally allowed regions.
Exceptions here are the pattern $C_{33}$ (the $T_1$-symmetry related pattern $C_{22}$) where $\t_y$ is bound in the inverted hierarchy type to be larger than $48^0$ (smaller
than $42^0$), the pattern $C_{31}$ (the $T_1$-symmetry related pattern $C_{21}$) in the normal hierarchy case where
 the Dirac angle $\d$ lies outside the interval $[123^0, 242^0]$ (approximately inside the interval $[60^0,300^0]$),
 and the pattern $C_{11}$ where the phases $\r$ and $\s$ tend to be far from $\pi/2$.

In all patterns, except $C_{11}$, there is a linear correlation in the inverted and degenerate cases between $\d$ and
$\r$ or $\s$, whence a sharp correlation of $J$ against these two latter phases. Also, a linear correlation exists
between $\r$ and $\s$ in all cases even though it ceases to be linear in the pattern $C_{11}$.

The limit $\t_z = 0$ can be attained in all patterns. In models ($C_{31}, C_{21}$ and
$C_{11}$), this limit corresponds to the singular inverted-hierarchy model with $m_3 = 0$. As to the parameter $m_{ee}$,
it can reach the value $0$, for the case of normal hierarchy, in the patterns $C_{33}, C_{22}$ and $C_{32}$, whereas this zero-limit can not be achieved in
the other patterns.

These features can help in distinguishing between the four independent models (let's take them as $C_{33},
C_{31}, C_{32}, C_{11}$). If the measured values of the mass ratios indicate an inverted hierarchy type then there are
only two acceptable models, out of the three models $C_{33},
C_{31}, C_{11}$ allowing for this type of hierarchy, depending on the intensity of this inverted hierarchy. If the intensity is strong ($m_2/m_3 > 10$)
 then the correct pattern would be either $C_{31}$ or $C_{11}$. A subsequent measurement of the phase angles ($\r$,$\s$) can exclude the pattern $C_{11}$ if it lies outside the `narrow' dotted region in plot l-I,L in Fig~\ref{c11fig1}.
 Now, if the intensity is mild($m_2/m_3 < 10$) then the choice would be between the patterns $C_{31}$ or $C_{33}$.
 In this case, a measurement of $\t_y$ smaller than $48^0$ would exclude the $C_{33}$ pattern, whereas, in case $\t_y > 48^0$,
 a measurement of the phase angles $\s,\d$ helps to decide which pattern can accommodate the data by comparing, say,
 to the `narrow' bands of plot e-I,L in Fig~\ref{c33fig1}. Table~\ref{signatures} summarizes these `experimental signatures' in the
 case of inverted hierarchy.

 \begin{table}[htbp]
\begin{center}
\begin{tabular}{|c||c|c|}
\hline
   Intensity & Mild & Strong \\
\hline
\hline
Acceptable patterns & $C_{33}$ \& $C_{13}$  & $C_{11}$ \& $C_{13}$ \\
\hline
signature & $\t_y < 48^0 \Rightarrow C_{13}$  &   \\
\hline
decisive phase angles & $\d,\s$  & $\r,\s$ \\
\hline
\end{tabular}
\end{center}
 \caption{\small  The inverted hierarchy `experimental' signatures distinguishing the different patterns.}
\label{signatures}
\end{table}

 If the mass measurements give a normal type of hierarchy, then the accepted patterns would be ($C_{32}, C_{33}$ and
 $C_{13}$). One indication here is when $\d$ lies inside $[123^0, 242^0]$ which would exclude $C_{31}$. The `narrow'
 band of the linear correlation between $\r$ and $\s$ in plot f-N,R in Fig~\ref{c31fig1} can help to exclude $C_{31}$
  when $\d$ lies outside the above interval. However, this measurement of $\r$ and $\s$ can not distinguish
  between $C_{32}$ and $C_{33}$ (look at the similar plots of f-N,R in Figs~\ref{c33fig1} and~\ref{c32fig1}), and it would be difficult to distinguish between these two patterns. For the degenerate type, the possible patterns would also be $C_{32}, C_{33}$ and
 $C_{13}$. Although we do not have a `signature' coming from the  $\d$ alone here, however the knowledge of all the phase
 angles together can distinguish between the patterns. This comes because the narrow `bands' corresponding to the linear correlations
 of ($\r, \s$) are different in plots f-D,R of Fig~\ref{c31fig1} and Fig~\ref{c32fig1} which would help to exclude the cases
 $C_{31}$ and $C_{32}$. Also, the different `band' structures of the linear correlations ($\d,\s$) in plots e-D,L of Fig~\ref{c33fig1} and Fig~\ref{c32fig1} would help to distinguish between the patterns $C_{33}$ and $C_{32}$.

Finally, all the models can be realized in the framework of flavor Abelian discrete
symmetry, with at most three additional SM-singlet scalar fields transforming appropriately,
implemented in seesaw schemes.

\section*{{\large \bf Acknowledgements}}
Both authors would like to thank  A. Smirnov and S. Petcov for useful discussions. Part of this work was done
within the associate schemes of ICTP and TWAS.

\end{document}